\documentclass[aps,prb,reprint,twocolumn,longbibliography,superscriptaddress]{revtex4-2}

\usepackage{graphicx}% Include figure files
\usepackage{bm}% bold math
\usepackage{bbold}
\usepackage{color}
\usepackage{hhline}
\usepackage{amsmath}
\usepackage{amssymb}

\def\br{ \bm{r} }
\def\bp{ \bm{p} }
\def\hbp{ \hat{\bm{p}} }
\def\bk{ \bm{k} }
\def\bq{ \bm{q} }

\def\Tr{ \mathrm{Tr} }
\def\tr{ \mathrm{tr} }
\def\hH{ {\hat H} }
\def\itO{ \mathit{\Omega} }
\def\hHb{ {\hat H}_{\text{band}} }

\def\hU{ {\hat{\mathsf{U}}} }
\def\hQ{ {\hat{\mathsf{Q}}} }
\def\hq{ {\hat{\mathsf{q}}} }
\def\Eg{ {\cal E}_g }
\def\ER{ {\cal E}_{\rm R} }
\def\aR{ {\alpha_{\rm R}} }
\def\hcU{ \hat{\cal U} }
\def\hcH{ \hat{\cal H} }
\def\hcJ{ \hat{\cal J} }
\def\hcT{ \hat{\cal T} }
\def\hcS{ \hat{\cal S} }
\def\hcO{ \hat{\cal O} }

\def\hcD{ \hat{\cal D} }
\def\hbs{ \hat{\bm{\sigma}} }

\def\msT{ \mathsf{T} }
\def\msS{ \mathsf{S} }
\def\msJ{ \mathsf{J} }
\def\msH{ \mathsf{H} }
\def\msG{ \mathsf{G} }
\def\msO{ \mathsf{O} }
\def\msW{ \mathsf{W} }
\def\bgam{ \bm{\gamma} }
\def\hbbone{ \hat{\mathbb{1}} }
\def\spup{ \lvert\uparrow\rangle }
\def\spdown{ \lvert\downarrow\rangle }

\begin{document}
\title{Spin currents in crystals with spin-orbit coupling:\\ multi-band effects in an effective Hamiltonian formalism}

\author{Kirill V. Samokhin}
\email{kirill.samokhin@brocku.ca}
\affiliation{Department of Physics, Brock University, St. Catharines, Ontario, Canada L2S 3A1}
\author{Manfred Sigrist}
\affiliation{Institute for Theoretical Physics, ETH Zurich, 8093 Zurich, Switzerland}
\author{Mark H. Fischer}
\affiliation{Department of Physics, University of Zurich, 8057 Zurich, Switzerland}

\date{\today}

\begin{abstract}
When focusing on a few essential bands in an effective description of a material to calculate observable quantities, the respective operators have to be adjusted accordingly. Ignoring contributions arising from integrating out remote bands can lead to qualitatively wrong results. 
We present a detailed analysis of the interband mixing effects on spin currents. Specifically, we calculate the intrinsic spin current in a time-reversal invariant noncentrosymmetric crystal in the presence of electron-lattice spin-orbit coupling. Starting from formally exact microscopic expressions, we derive the spin current operator restricted to one or more essential bands by iterative elimination of the contributions from distant bands. 
We show that the standard definition of the spin current operator
in terms of the group velocity obtained from an effective band Hamiltonian cannot be justified using a microscopic theory. The modified expression for the spin current operator contains additional terms, which dominate the equilibrium spin current in a uniform crystal. We show that the magnitude of these additional terms can considerably exceed the spin current obtained using the standard definition. 
\end{abstract}

\maketitle
   
%==================================================================
%==================================================================

\section{Introduction}
\label{sec: introduction}

A spin current is the flow of spin angular momentum of electrons, which can exist either with or without an accompanying flow of electron charge. Spin currents are fundamentally important to modern spintronics applications, in particular to semiconductor spintronic devices, in which injection of nonequilibrium spins and manipulation of the spin polarization is achieved using the electron spin-orbit coupling (SOC)~\cite{spintronics-review-1,spintronics-review-2,spintronics-review-3}.
In general, we distinguish two types of spin currents. For spintronics applications, \textit{extrinsic} spin currents driven by an applied electric field, as in the spin Hall effect~\cite{DP71,Hir-PRL99,Murakami03,Sinova04,ERH07,SHE-review-2015}, are usually considered. 
Additionally, \textit{intrinsic} spin currents are present even in equilibrium in materials with gyrotropic point-group symmetry~\cite{Rashba03, DRRT22}. 
These latter spin currents are dissipationless and, as such, are not directly related to spin transport and spin accumulation near inhomogeneities, such as the sample boundaries.

Experimental observability of the equilibrium spin currents has been a somewhat controversial subject. It was argued in Ref. \onlinecite{Rashba03} that these currents ``do not describe any real transport of electron spins and cannot result in spin injection or accumulation'', which would make their detection difficult if possible at all. However, according to Refs. \onlinecite{Sonin07,Sonin10}, the equilibrium spin current in a nonuniform system is accompanied by a nonzero spin torque, so that spins are transported without accumulation between the regions with opposite signs of the torque, which serve as spin sources or sinks. Near the sample boundaries, spin torque can be detected by measuring the mechanical torque exerted by the sample on the substrate~\cite{Sonin-PRL07}. 
Another proposal to measure the equilibrium spin currents utilizes the fact that they induce an electric field~\cite{Hir99,Sun04}, which can potentially be probed in experiment.

How to calculate the spin current carried by electrons in a crystal lattice? For time-reversal (TR) invariant systems without SOC, the electron bands are twofold degenerate due to spin, regardless of the presence of an inversion center, and the current of the $\mu$th component of spin in the $i$th direction is described by the operator $\hat J_{\mu,i}(\bk)=v_i(\bk)\hat\sigma_\mu$, where $\bm{v}(\bk)=\hbar^{-1}\partial\xi/\partial\bk$ is the group velocity of quasiparticles in the band with dispersion $\xi(\bk)$ and $\hbs=(\hat{\sigma}_x,\hat{\sigma}_y,\hat{\sigma}_z)$ are the spin Pauli matrices \cite{spin-current-units}. As the spin current is diagonal in band space, the total spin current can be simply calculated by summing over bands. Note that the above expression is a direct generalization of the charge current operator $\hat{\bm{J}}^{\text{c}}(\bk)=-e\bm{v}(\bk)\hat\sigma_0$, where $-e$ is the electron charge and $\hat\sigma_0$ is the identity matrix in the spin space.

The situation is more complicated if we take SOC of electrons with the crystal lattice into account. The first and most obvious complication is that spin is no longer a good quantum number and the total spin is not a conserved quantity, which leads to ambiguity in the microscopic definitions of the spin current and the spin torque~\cite{Sonin07}. 
Still, we can use a non-Abelian gauge invariance of the microscopic Pauli Hamiltonian to define the microscopic spin current operator, as discussed in Refs.~\onlinecite{Tokatly08,DRRT22}.

However, even after settling on the microscopic definition of the spin current, for practical calculations, such as those of the equilibrium spin current or the spin Hall conductivity, an explicit expression for the spin current operator restricted to one or more essential bands is desirable. In particular, we would like to calculate the spin current within an effective band Hamiltonian $\hcH(\bk)$, such as the Rashba model. It is this second issue that we focus on in this paper. 

We can derive an effective Hamiltonian microscopically using $\bk\cdot\bp$ theory~\cite{Winkler-book,kp-book}, in which the contributions of distant bands---all bands other than the essential ones---are ``integrated out'' by a partial diagonalization of the complete microscopic Hamiltonian. While the $\bk\cdot\bp$ method has been extensively used to obtain effective Hamiltonians for many different materials since the 1950s, calculations of the spin current within the same models have received considerably less attention. Importantly, it is well known that SOC in noncentrosymmetric materials mostly derives from inter-orbital mixing~\cite{Kane57,RS59,LYV96,Peter00}. As such, the question arises how eliminating distant bands affects the definition of the spin current in an effective Hamiltonian formalism.
Similar effects of interband transitions on transport properties have attracted a lot of attention recently in the context of what is known as quantum geometry~\cite{Verma25,Yu25}.

\begin{figure}[t]
    \includegraphics{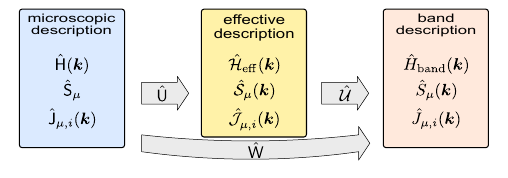}
    \caption{Outline of our calculation of the spin current: Starting from a full microscopic description, we can either try to directly calculate observables in the band representation, or first derive effective Hamiltonians in the much smaller essential orbital subspaces. In this latter description, one needs to find the correct form of the spin current operator $\mathcal{J}_{\mu,i}(\bk)$.}
    \label{fig:overview}
\end{figure}

The paper is organized as follows. After a short discussion of the simple Rashba model in the next section, in Sec. \ref{sec: exact microscopic} we outline the microscopic formalism for spin-related observables in crystals with SOC. This formalism, while exact for non-interacting electrons, involves dealing with infinite matrices and is therefore not suitable for practical calculations.
In Sec. \ref{sec: effective H formalism}, we develop an effective Hamiltonian description of the spin observables by focusing on a small number of essential bands and eliminating the interband couplings by a canonical transformation. This results in an expression for the spin current operator which significantly differs from the commonly used one. Figure~\ref{fig:overview} schematically shows our program and summarizes the notations for the observable quantities in different bases.
As an example, the general theory is applied to a quasi-2D noncentrosymmetric metal of tetragonal symmetry in Sec. \ref{sec: applications-C4v}. Most of the technical details are banished to the Appendices.

%==================================================================
%==================================================================

\section{Intrinsic spin current in a single band model}
\label{sec: intrinsic SC}

Before introducing the general formulation, we recap some known results for a noncentrosymmetric crystal in the absence of intrinsic magnetic order and any external fields. The simplest effective Hamiltonian describing one band split by SOC has the form
\begin{equation}
\label{general-Rashba-model}
	\hcH(\bk)=\varepsilon(\bk)\hat\sigma_0+\bgam(\bk)\cdot\hbs,
\end{equation}
where the second term with $\bgam(\bk)=-\bgam(-\bk)$ is an antisymmetric SOC. 
Diagonalizing Eq.~(\ref{general-Rashba-model}), one obtains two bands $\xi_\lambda(\bk)=\varepsilon(\bk)+\lambda|\bgam(\bk)|$, which are labelled by the ``helicity'' index $\lambda=\pm$. 
While the well-known Rashba model corresponds to $\bgam(\bk)=\alpha_{\rm R}(k_y,-k_x,0)$~\cite{Rashba-model,Manchon15}, different symmetries of the system can impose a more complicated $\bk$-dependence~\cite{Dressel55,Smidman17,Sam19}. 
If the essential physics of the system is determined by more than one band, then $\hcH(\bk)$ is represented by a larger matrix, whose elements in the vicinity of the $\Gamma$ point (or any other symmetry point in the Brillouin zone) are polynomial functions of $\bk$. Phenomenologically, these functions are constrained by symmetry and can be obtained, for instance, using the method of invariants~\cite{Bir-Pikus-book}. 

A commonly used expression for the spin current was introduced in Refs.~\onlinecite{Rashba03,Sinova04}. From the effective Hamiltonian $\hcH(\bk)$, we obtain the group velocity operator $\hat{\bm{v}}=\hbar^{-1}\partial\hcH/\partial\bk$ and then define the spin current operator as
\begin{equation}
\label{spin current-vertex-Rashba}
	\hcJ_{\mu,i}(\bk)=\frac{1}{2}\biggl\{\frac{1}{\hbar}\frac{\partial\hcH(\bk)}{\partial k_i},\hat\sigma_\mu\biggr\},
\end{equation}
where the curly brackets denote the anticommutator. The average spin current in a uniform equilibrium state at temperature $T$ is given by
\begin{equation}
\label{j-mu-i-average-lambdas}
  \langle j_{\mu,i}\rangle = \sum_{\lambda=\pm}\int\frac{d^2\bk}{(2\pi)^2}\, \langle \bk,\lambda|\hcJ_{\mu,i}(\bk)|\bk,\lambda\rangle f(\xi_\lambda),
\end{equation}
where $|\bk,\lambda\rangle$ are the eigenstates of $\hcH(\bk)$, $f(\xi)=[e^{(\xi-\mu)/T}+1]^{-1}$ is the Fermi function, and $\mu$ is the chemical potential. Here and in the following, we assume a two-dimensional (2D) 
geometry with $\bk=(k_x,k_y)$, but our results can be straightforwardly generalized to three dimensions.

While the spin current vanishes in centrosymmetric crystals by symmetry, it can be nonzero in a noncentrosymmetric crystal, even in thermodynamic equilibrium. In the model described by Eq.~(\ref{general-Rashba-model}) with $\varepsilon(\bk)=\hbar^2k^2/2m^*$ ($m^*$ is the effective mass) and $\bgam(\bk)=\alpha_{\rm R}(k_y,-k_x,0)$, we obtain from the definition in Eq.~(\ref{spin current-vertex-Rashba}) that the nonzero components of the equilibrium spin current at $T=0$ and $\mu>0$ are given by~\cite{Rashba03}
\begin{equation}
\label{j-xy-Rashba}
  \langle j_{x,y}\rangle=-\langle j_{y,x}\rangle=J_s(T=0)=\frac{m^{*,2}}{3\pi\hbar^5}\alpha_{\rm R}^3.
\end{equation}
Surprisingly, the result does not depend on the chemical potential, in other words, the electron concentration. We also note that the equilibrium bulk spin current is cubic in the Rashba SOC strength. 

There are two different ways to obtain the Rashba term in the effective Hamiltonian, Eq.~(\ref{general-Rashba-model}), from microscopic theory. Treating the electron-lattice SOC as a small perturbation, this term arises either as the first-order correction or as a second-order correction containing the usual $\bk\cdot\bp$ perturbation and the $\bk$-independent atomic-like SOC~\cite{Kane57,RS59}. The former mechanism does not require any interband transitions and produces the ``intrinsic'' Rashba SOC even in an isolated band, which is certainly allowed by symmetry. However, its magnitude turns out to be much smaller than that of the second contribution, which is determined by the interband mixing~\cite{Kane57,RS59,LYV96,Peter00}. Thus, a proper quantitative treatment of the effective SOC and the spin current must include multi-band or multi-orbital effects.

%==================================================================
%==================================================================

\section{Microscopic formalism}
\label{sec: exact microscopic}

\subsection{Kohn-Luttinger basis}

We consider non-interacting electrons in a crystal described by the point group $\mathbb{G}$ and with conserved TR symmetry. To develop a convenient microscopic representation of the Hamiltonian, we start with the Luttinger-Kohn (LK) basis~\cite{LK55}, see Appendix~\ref{app: k-p-theory},
\begin{equation}
\label{LK-basis}
  |\bk,\ell, q \alpha\rangle=\frac{1}{\sqrt{\cal V}}e^{i\bk\br}f_{\ell q}(\br)|\alpha\rangle,
\end{equation}
where $\bk$ is the wave vector in the first Brillouin zone (BZ), ${\cal V}$ is the volume of the system, and $|\alpha\rangle = \spup,\spdown$ are the basis spinors. 
The lattice-periodic functions $f_{\ell q}(\br)$ are the eigenstates of the reduced Hamiltonian at the $\Gamma$ point including the crystal field but excluding SOC. The corresponding energy eigenvalues are denoted by $\epsilon_\ell$. The functions $f_{\ell q}(\br)$, called the orbital states or simply the orbitals, transform according to a $d_\ell$-dimensional single-valued irreducible representation (irrep) $\gamma_\ell$ of the crystal point group $\mathbb{G}$, with the additional index $q=1,...,d_\ell$ labelling the orbital degeneracy. We assume a strong crystal field, so that the orbital states are different from the atomic orbitals, in particular, they do not have a definite parity in a noncentrosymmetric crystal.

One can calculate the matrix elements of various physical observables in the LK basis, Eq.~(\ref{LK-basis}), see Appendices \ref{app: k-p-theory} and \ref{app: spin current-LK} for details. 
The complete microscopic Hamiltonian is represented by the matrix $\hat\msH(\bk)=\hat{\msH}'(\bk)+\hat{\msH}''(\bk)$, where
\begin{equation}
\label{H-exact}
  \hat{\msH}'= \begin{pmatrix}
       \hat h_1 & 0 & \cdots \\
       0 & \hat h_2 & \cdots \\
       \vdots & \vdots & \ddots
       \end{pmatrix},\quad
       \hat{\msH}''= \begin{pmatrix}
       0 & \hat h_{12} & \cdots \\
       \hat h_{21} & 0 & \cdots \\
       \vdots & \vdots & \ddots
       \end{pmatrix}.
\end{equation}
All intra-orbital contributions, including the intra-orbital SOC terms, are collected into the $2d_\ell\times 2d_\ell$ matrices $\hat h_\ell(\bk)$, while $\hat h_{\ell\ell'}(\bk)=\hat h^\dagger_{\ell'\ell}(\bk)$ is a $2d_\ell\times 2d_{\ell'}$ matrix describing the coupling between the states $\ell$ and $\ell'$. The matrix elements of $\hat{\msH}'$ and $\hat{\msH}''$ are analytic functions of $\bk$, namely zeroth, first, or second degree polynomials, whose form depends on the symmetry of the orbital states (see an example in Appendix \ref{app: p-p-example}).

The spin operator in the LK basis has the form
\begin{equation}
\label{S-mu-exact}
	\hat{\msS}_\mu=\left( \begin{array}{ccc}
                  \hbbone_1\otimes\hat\sigma_\mu & 0 & \cdots \\
                  0 & \hbbone_2\otimes\hat\sigma_\mu & \cdots \\
                  \vdots & \vdots & \ddots
                  \end{array} \right),
\end{equation}
where $\hbbone_\ell$ is the $d_\ell\times d_\ell$ identity matrix in the $\ell$th orbital subspace. The microscopic spin-current operator is represented by the matrix
\begin{equation}
\label{J-mu-i-exact}
  \hat{\msJ}_{\mu,i}(\bk)=\frac{1}{2}\biggl\{\frac{1}{\hbar}\frac{\partial\hat{\msH}(\bk)}{\partial k_i},\hat{\msS}_\mu\biggr\},
\end{equation}
whereas for the spin-torque operator, we have
\begin{equation}
\label{T-mu-exact}
  \hat{\msT}_\mu(\bk)=\frac{i}{\hbar}\bigl[\hat{\msH}(\bk),\hat{\msS}_\mu\bigr],
\end{equation}
which is nonzero only in the presence of SOC. For comparison, the charge-current operator is given by
\begin{equation}
\label{charge-J-exact}
  \hat{\msJ}^{\text{c}}_{i}(\bk)=-\frac{e}{\hbar}\frac{\partial\hat{\msH}(\bk)}{\partial k_i}.
\end{equation}

We emphasize that the expressions in Eqs.~(\ref{S-mu-exact})-(\ref{charge-J-exact}) are \textit{exact} for non-interacting electrons, if (i) variation of the observable quantities on the scale of the lattice constant is averaged and (ii) all orbital states at the $\Gamma$ point are taken into account. If one includes $N$ orbital states in total (i.e. $\sum_\ell d_\ell=N$, with $N\to\infty$ in a complete basis), then all observables are represented by $2N\times 2N$ matrices.

\subsection{Band representation}
\label{sec: observables-band-rep}

The band structure is obtained by diagonalizing the microscopic Hamiltonian matrix, $\hat{\msH}(\bk)|\bk,n\rangle=\xi_n(\bk)|\bk,n\rangle$. 
Although in the presence of SOC spin is no longer a good quantum number, in centrosymmetric crystals the electron bands remain twofold degenerate at each wave vector $\bk$, due to the combined symmetry $KI$, where $K$ is the TR operation and $I$ is the spatial inversion~\cite{Kittel-book}.
In contrast, in noncentrosymmetric crystals the band degeneracy is lifted at a generic wave vector. Diagonalizing the Hamiltonian, we have
\begin{align}
\label{WHW}
  \hHb(\bk) & = \hat{\msW}^{-1}(\bk)\hat{\msH}(\bk)\hat{\msW}(\bk)\nonumber\\
  & = \begin{pmatrix}
      \xi_1(\bk) & 0 & \cdots & 0 \\
      0 & \xi_2(\bk) & \cdots & 0 \\
      \vdots & \vdots & \ddots & \vdots \\
      0 & 0 & \cdots & \xi_{2N}(\bk)
      \end{pmatrix},
\end{align}
where $\hat{\msW}$ is a $2N\times 2N$ unitary matrix. Due to TR symmetry, the band dispersions satisfy $\xi_n(\bk)=\xi_n(-\bk)$.

We refer to the description of the system in the LK basis as the \textit{microscopic} or \textit{orbital} representation, whereas the transformation (\ref{WHW}) defines the \textit{band} representation. 
In Sec. \ref{sec: effective H formalism} below, we introduce another, intermediate representation in terms of the \textit{effective} Hamiltonian $\hcH_{\text{eff}}(\bk)$, which is obtained by eliminating the contributions of non-essential orbital states from the microscopic Hamiltonian $\hat{\msH}(\bk)$.

The matrix $\hat{\msW}$ can be used to transform other observables $\hat{\msO}(\bk)$ from the orbital into the band representation, $\hat O(\bk)=\hat{\msW}^{-1}(\bk)\hat{\msO}(\bk)\hat{\msW}(\bk)$, which facilitates the calculation of their expectation values. For instance, the transformed spin-torque operator $\hat T_{\mu}(\bk)=(i/\hbar)[\hHb(\bk),\hat S_\mu(\bk)]$,
where $\hat S_\mu(\bk)=\hat{\msW}^{-1}(\bk)\hat{\msS}_\mu\hat{\msW}(\bk)$, has vanishing diagonal elements and therefore its average in a uniform equilibrium state is equal to zero.

For the observables involving derivatives of the Hamiltonian, the situation is more interesting. Let us start with the charge current, see Eq. (\ref{charge-J-exact}), whose matrix elements in the band basis are given by
\begin{eqnarray}
\label{charge-current-operator-band}
    J^{\text{c}}_{i,mn}(\bk) &=& \bigl[\hat{\msW}^{-1}(\bk)\hat{\msJ}^{\text{c}}_i(\bk)\hat{\msW}(\bk)\bigr]_{mn}\nonumber\\
    &=& -ev_{n,i}\delta_{mn}+i\frac{e}{\hbar}(\xi_m-\xi_n)\itO^i_{mn},
\end{eqnarray}
where $\bm{v}_n=\hbar^{-1}\bm{\nabla}_{\bk}\xi_n$ is the quasiparticle velocity in the $n$th band and $\hat{\bm{\itO}}=-i\hat{\msW}^{-1}\bm{\nabla}_{\bk}\hat{\msW}$ is the Berry connection matrix. The elements of the latter are given by $\bm{\itO}_{mn}(\bk)=-i\langle\varphi_{\bk,m}|\bm{\nabla}_{\bk}|\varphi_{\bk,n}\rangle$, where $\varphi_{\bk,n}$ are the lattice-periodic parts of the Bloch spinor wave functions. We note that the Berry connections, which encode the quantum geometry of the Bloch bands, enter only the off-diagonal elements of the charge current, Eq.~(\ref{charge-current-operator-band}). Therefore, the quantum-geometrical effects due to interband transitions can affect only the properties determined by the \textit{correlators} of the charge-current operator. One prominent example is the existence of a lower bound on the superfluid stiffness, which is extracted from the current-current response function in the superconducting state~\cite{Peotta15,Liang17}.     

From Eq.~(\ref{J-mu-i-exact}), we obtain the band representation  of the spin current
\begin{equation}
\label{spin-current-operator-band}
    \hat J_{\mu,i}(\bk)=\frac{1}{2\hbar}\biggl\{\frac{\partial\hHb}{\partial k_i}+i[\hat{\itO}_i,\hHb],\hat S_\mu\biggr\},
\end{equation}
with the diagonal matrix elements
\begin{eqnarray}
\label{spin-J-intraband}
    && J_{\mu,i}^{nn}(\bk) = v_{n,i}S_\mu^{nn} \nonumber\\
    && \qquad +\frac{i}{2\hbar}\sum_{m\neq n}(\xi_m-\xi_n)(\itO^i_{nm}S_\mu^{mn}-\itO^i_{mn}S_\mu^{nm}).
\end{eqnarray}
One can see that the spin current carried by quasiparticles in the $n$th band is not just equal to a product of their group velocity and the band-transformed spin. There is an additional term given by the second line in Eq. (\ref{spin-J-intraband}), which does not admit a simple interpretation. Notably, in contrast to the charge current, the interband Berry connections appear already in the diagonal elements of $\hat J_{\mu,i}(\bk)$ and are therefore expected to affect even the \textit{equilibrium} spin current: 
\begin{equation}
\label{j-mu-i-average-bands}
  \langle j_{\mu,i}\rangle = \sum_{n}\int\frac{d^2\bk}{(2\pi)^2}\, J_{\mu,i}^{nn}(\bk) f(\xi_n).
\end{equation}
To get a sense of the relative importance of the ``trivial'' and ``topological'' contributions to the spin current, given by the first and second lines in Eq. (\ref{spin-J-intraband}) respectively, one can evaluate them in some simple band structure models. An explicit calculation in one such model in Appendix \ref{app: trivial-topological-toy} shows that the two contributions are very similar in structure and magnitude.  

The expression in Eq.~(\ref{spin-J-intraband}) indicates that the effects of interband transitions, which are encoded in both $\hat{\bm{\itO}}(\bk)$ and $\hat S_\mu(\bk)$, play an essential role in spin transport. However, using the full microscopic formalism as a starting point is clearly impractical, as one has to deal with large matrices, Eqs. (\ref{H-exact}) and (\ref{J-mu-i-exact}). Below we develop a formalism that allows an explicit analytical calculation of the multi-band effects.

%==================================================================
%==================================================================

\section{Effective Hamiltonian formalism}
\label{sec: effective H formalism}

\subsection{General case}

To calculate the spin current in the presence of SOC, we shall use the following procedure: First, we identify the orbital state or states with energy close to the chemical potential, which dominate the low-energy physics of the system. We call these orbitals \textit{essential}, whereas all other orbitals are \textit{distant}. 
The properties of the essential orbitals are described by an effective Hamiltonian, in which the couplings with the distant ones have been eliminated perturbatively. Next, using the same perturbation theory, we derive expressions for other spin observables, such as the spin current, in the essential orbital subspace. 
Finally, we diagonalize the effective Hamiltonian to find the band structure as well as the average spin current in a uniform equilibrium state. The key question we aim to answer is the following: Does the spin current operator obtained by this procedure have the form of Eq.~(\ref{spin current-vertex-Rashba})?

To eliminate the inter-orbital couplings and bring the Hamiltonian to a block-diagonal form, we apply a canonical transformation---known as Luttinger-Kohn~\cite{LK55} or Schrieffer-Wolff~\cite{SW66} transformation---to Eq.~(\ref{H-exact}),
\begin{eqnarray}
\label{LK-transform}
    \hcH_{\rm eff}(\bk) &\equiv& \hU^{-1}(\bk)\hat{\msH}(\bk)\hU(\bk) \nonumber\\
       &=& \begin{pmatrix}
       \hcH_1(\bk) & 0 & \cdots \\
       0 & \hcH_2(\bk) & \cdots \\
       \vdots & \vdots & \ddots
       \end{pmatrix},
\end{eqnarray}
where the $2d_\ell\times 2d_\ell$ matrix $\hcH_\ell(\bk)$ is interpreted as the effective Hamiltonian in the $\ell$th orbital subspace. The unitary transformation matrix is given by $\hU=\exp(i\hQ)$, where
\begin{equation}
\label{Q-matrix}
      \hQ(\bk)=\begin{pmatrix}
       0 & \hQ_{12}(\bk) & \cdots \\
       \hQ_{21}(\bk) & 0 & \cdots \\
       \vdots & \vdots & \ddots
       \end{pmatrix}
\end{equation}
is a Hermitian matrix with only inter-orbital elements. Details of the procedure are presented in Appendix \ref{app: derivation of H-eff}.

The effective Hamiltonian can be obtained perturbatively to any desired order in the inter-orbital couplings. For instance, the second-order result reads 
\begin{eqnarray}
\label{H-eff-2nd-order-final}
    && \hcH_\ell(\bk) = \hat h_\ell(\bk) \nonumber\\
    && \qquad +\frac{i}{2}\sum_{\ell'\neq\ell}\bigl[\hat h_{\ell\ell'}(\bk)\hQ^\dagger_{\ell\ell'}(\bk)-\hQ_{\ell\ell'}(\bk)\hat h^\dagger_{\ell\ell'}(\bk)\bigr],
\end{eqnarray}
where the $2d_{\ell}\times 2d_{\ell'}$ matrix $\hQ_{\ell\ell'}$ is determined by 
\begin{equation}
\label{Q-1st-order-final}
	\hat h_{\ell}(\bk)\hQ_{\ell\ell'}(\bk)-\hQ_{\ell\ell'}(\bk)\hat h_{\ell'}(\bk)=i\,\hat h_{\ell\ell'}(\bk).
\end{equation}
If the spacings $\epsilon_{\ell} - \epsilon_{\ell'}$ at the $\Gamma$ point between different orbital states $\ell$ and $\ell'$ are much larger than all the intra- and inter-orbital energy scales, we can replace $\hat h_{\ell}$ by $\epsilon_\ell$ and obtain the estimate $\hQ_{\ell\ell'}(\bk)\sim i\hat h_{\ell\ell'}(\bk)/(\epsilon_\ell-\epsilon_{\ell'})$ to leading order. In some cases, Eq.~(\ref{Q-1st-order-final}) can be solved without making this assumption, as we will demonstrate in Sec.~\ref{sec: applications-C4v}.

The transformation given by Eq.~(\ref{LK-transform}) also affects other observables, such as the spin current and the spin torque, which do not have a block-diagonal form, in general. From Eqs.~(\ref{J-mu-i-exact}) and (\ref{T-mu-exact}) we obtain
\begin{eqnarray}
\label{spin-current-LK-transformed}
	\hat{\cal J}_{\mu,i}(\bk) &=& \hU^{-1}(\bk) \hat{\msJ}_{\mu,i}(\bk) \hU(\bk) \nonumber\\
	&=& \frac{1}{2}\biggl\{\frac{1}{\hbar}\frac{\partial\hcH_{\rm eff}}{\partial k_i}+\frac{i}{\hbar}[\hat{\mathsf{\Omega}}_{i},\hcH_{\rm eff}],\hat{\cal S}_\mu\biggr\}
\end{eqnarray}
and 
\begin{equation}
\label{spin-torque-LK-transformed}
	\hcT_\mu(\bk)=\hU^{-1}(\bk) \hat{\msT}_\mu(\bk) \hU(\bk)=\frac{i}{\hbar}\bigl[\hcH_{\text{eff}},\hat{\cal S}_\mu\bigr],
\end{equation}
where $\hat{\bm{\mathsf{\Omega}}}(\bk)=-i\,\hU^{-1}\bm{\nabla}_{\bk}\hU$ is reminiscent of the Berry connection introduced in Sec. \ref{sec: observables-band-rep} and
\begin{equation}
\label{spin-operator-LK-transformed}
	\hcS_\mu(\bk)=\hU^{-1}(\bk)\hat{\msS}_\mu\hU(\bk)
\end{equation}
is the transformed spin operator. All matrices here are analytic functions of $\bk$ by construction and have both intra-orbital and inter-orbital blocks,
\begin{equation}
\label{SJT-blocks}
  \hcO(\bk) = \begin{pmatrix}
       \hcO_{11}(\bk) & \hcO_{12}(\bk) & \cdots \\
       \hcO_{21}(\bk) & \hcO_{22}(\bk) & \cdots \\
       \vdots & \vdots & \ddots
       \end{pmatrix}
\end{equation}
with $\hcO(\bk)=\hcJ_{\mu,i}(\bk)$, $\hcT_\mu(\bk)$, or $\hcS_\mu(\bk)$. Each diagonal block is a $2d_\ell\times 2d_\ell$ matrix, which can be interpreted as the \textit{effective observable} in the $\ell$th orbital subspace. 

Treating the inter-orbital couplings as a perturbation, we can expand in powers of $\hQ$,
\begin{equation}
  \hat{\bm{\mathsf{\Omega}}}=\bm{\nabla}_{\bk}\hQ-\frac{i}{2}[\hQ,\bm{\nabla}_{\bk}\hQ]+\ldots
\end{equation}
and 
\begin{equation}
  \hat{\cal S}_{\mu}=\hat{\msS}_\mu-i[\hQ,\hat{\msS}_\mu]-\frac{1}{2}[\hQ,[\hQ,\hat{\msS}_\mu]]+\ldots.
\end{equation}
Substituting these expansions in Eq.~(\ref{spin-current-LK-transformed}) and using Eq.~(\ref{H-eff-2nd-order-final}), we obtain the effective spin-current operator to second order in $\hQ$,
\begin{eqnarray}
\label{spin current-projected-final}
  \hcJ_{\mu,i}^{\ell\ell} &=& \frac{1}{2}\biggl\{\frac{1}{\hbar}\frac{\partial\hcH_\ell}{\partial k_i},\hat{\mathbb{1}}_\ell\otimes\hat\sigma_\mu\biggr\} \nonumber\\
    && -\frac{1}{4}\biggl\{\frac{1}{\hbar}\frac{\partial\hat{\msH}'}{\partial k_i},[\hQ,[\hQ,\hat{\msS}_\mu]]\biggr\}_\ell \nonumber\\
    && +\frac{1}{2\hbar}\biggl\{\biggl[\frac{\partial\hQ}{\partial k_i},\hat{\msH}'\biggr],[\hQ,\hat{\msS}_\mu]\biggr\}_\ell \nonumber\\
    && +\frac{1}{4\hbar}\biggl\{\biggl[\biggl[\hQ,\frac{\partial\hQ}{\partial k_i}\biggr],\hat{\msH}'\biggr],\hat{\msS}_\mu\biggr\}_\ell.
\end{eqnarray}
Here, the subscripts $\ell$ indicates the projection onto the $\ell$th orbital subspace. Analogously, we calculate the effective spin torque ($\hcT_{\mu}^{\ell\ell}$) and spin ($\hcS_{\mu}^{\ell\ell}$) operators. Corrections to first order in $\hQ$ appear only 
through the inter-orbital blocks of Eq.~(\ref{SJT-blocks}) and, therefore, do not contribute to the effective intra-orbital operators. 
Obviously, only the first term of Eq.~(\ref{spin current-projected-final}) corresponds to the form of the spin-current operator introduced in Eq.~(\ref{spin current-vertex-Rashba}) and the other terms are important corrections of second order in $\hQ$. 

Finally, the band structure is found by diagonalizing the effective Hamiltonian, $\hHb(\bk)=\hcU^{-1}(\bk)\hcH_{\rm eff}(\bk)\hcU(\bk)$, where 
$$
    \hcU(\bk)=\begin{pmatrix}
       \hcU_1(\bk) & 0 & \cdots \\
       0 & \hcU_2(\bk) & \cdots \\
       \vdots & \vdots & \ddots
       \end{pmatrix}.
$$
Each diagonal block $\hcU_\ell(\bk)$ is a $2d_\ell\times 2d_\ell$ unitary matrix in the $\ell$th orbital subspace. From Eqs.~(\ref{WHW}) and (\ref{LK-transform}), we have $\hat{\msW}(\bk)=\hU(\bk)\hcU(\bk)$. 
The same transformation also produces the observables in the band representation: $\hat{O}(\bk)=\hcU^{-1}(\bk)\hcO(\bk)\hcU(\bk)$. Note that, while $\hU$ is by construction analytic in $\bk$, $\hcU$ and $\hat{\msW}$ are not necessarily so.

\subsection{Focus on one essential orbital}
\label{sec: one-orbital}

Assuming that only one orbital state is close to the chemical potential and dropping the index $\ell$, our findings can be summarized as follows. The effective Hamiltonian is given by a $2d\times 2d$ matrix $\hcH(\bk)$, where $d$ is the dimension of the essential-orbital subspace. The effective spin-current operator has the form
\begin{equation}
\label{spin current-eff-general}
	\hcJ_{\mu,i}(\bk)=\frac{1}{2}\biggl\{\frac{1}{\hbar}\frac{\partial\hcH(\bk)}{\partial k_i},\hat{\mathbb{1}}\otimes\hat\sigma_\mu\biggr\}+\delta\hcJ_{\mu,i}(\bk),
\end{equation} 
where $\hat{\mathbb{1}}$ is the $d\times d$ identity matrix. The first term reproduces the standard definition, Eq.~(\ref{spin current-vertex-Rashba}). 
The second term, which comprises the second, third, and fourth terms in Eq.~(\ref{spin current-projected-final}), \textit{cannot} be expressed in terms of the effective Hamiltonian alone. 
We also note that, according to the second line of Eq.~(\ref{spin-current-LK-transformed}), the effects of the inter-orbital couplings on $\hcJ_{\mu,i}$ cannot be absorbed into the renormalization of the spin operator, in other words $\hat{\cal J}_{\mu,i}\neq(1/2\hbar)\{\partial\hcH/\partial k_i,\hat{\cal S}_\mu\}$. In contrast, the spin torque
\begin{equation}
\label{spin-torque-eff-general}
	\hat{\cal T}_\mu(\bk)=\frac{i}{\hbar}\bigl[\hcH(\bk),\hat{\cal S}_\mu(\bk)\bigr]
\end{equation}
is entirely determined by the effective Hamiltonian and the effective spin operator.

Before continuing, we note that the expression in Eq.~(\ref{spin current-eff-general}) is reduced to Eq.~(\ref{spin current-vertex-Rashba}) only if the mixing of the essential orbital state with all other states is completely neglected, which corresponds to the limit of \textit{infinitely distant} orbitals, when $\hQ=0$ and $\delta\hcJ_{\mu,i}=0$. 
However, in this limit the effective Hamiltonian $\hcH(\bk)$ is necessarily the same as the ``bare'' intra-orbital Hamiltonian $\hat h(\bk)$, according to Eq.~(\ref{H-eff-2nd-order-final}). 
The simplest nontrivial model must therefore include some coupling with at least one distant orbital state, in which case $\hcH$ differs from $\hat h$ and $\delta\hcJ_{\mu,i}\neq 0$. One such model, in which Eqs.~(\ref{H-eff-2nd-order-final}) and (\ref{spin current-projected-final}) can be evaluated analytically in a closed form, is discussed in the next section. 

Next, we calculate the electron band structure by diagonalizing the effective single-orbital Hamiltonian, $\hcH(\bk)|\bk,n\rangle=\xi_n(\bk)|\bk,n\rangle$, where $n=1,...,2d$. 
For the expectation values of observables in a uniform equilibrium state, we obtain
\begin{equation}
\label{SJT-average} 
  \langle O\rangle = \sum_{n}\int\frac{d^2\bk}{(2\pi)^2}\, \langle\bk,n|\hcO(\bk)|\bk,n\rangle f(\xi_n),
\end{equation}
as explained in Appendices \ref{app: spin current-LK} and \ref{app: derivation of H-eff}. 

Some observables average to zero due to the TR invariance requirement. Indeed, one can show that the effective Hamiltonian satisfies the constraint (see Appendix \ref{app: TRI})
\begin{equation}
\label{H-eff-TR-constraint}
  \hcD^{-1}(K)\hcH(\bk)\hcD(K)=\hcH^*(-\bk),\quad 
\end{equation}
where $\hcD(K)=\hat{\mathbb{1}}\otimes(-i\hat\sigma_y)$ is the matrix representation of the TR operation in the essential orbital subspace. More generally, we find
\begin{equation}
\label{SJT-eff-TR-constraint}
  \hcD^{-1}(K)\hcO(\bk)\hcD(K)=\pm\hcO^*(-\bk),
\end{equation}
with the plus sign for TR-even observables, $\hcO(\bk)=\hcJ_{\mu,i}(\bk)$ or $\hcT_{\mu}(\bk)$, and the minus sign for TR-odd observables, $\hcO(\bk)=\hcS_{\mu}(\bk)$. 
In a noncentrosymmetric crystal, the bands are nondegenerate almost everywhere in the BZ and the states $[\hcD^{-1}(K)|\bk,n\rangle]^*$ and $|-\bk,n\rangle$ can only differ by a phase factor, according to the TR invariance constraint of Eq.~(\ref{H-eff-TR-constraint}). Therefore, the expectation values satisfy
\begin{equation}
\label{SJT-expectation-values-TR}
  \langle\bk,n|\hcO(\bk)|\bk,n\rangle=\pm\langle -\bk,n|\hcO(-\bk)|-\bk,n\rangle,
\end{equation}
with the plus (minus) sign for the TR-even (odd) observables. To obtain this result, we used the Hermiticity of the matrix $\hcO(\bk)$ and Eq. (\ref{SJT-eff-TR-constraint}).

It follows from Eq.~(\ref{SJT-expectation-values-TR}) and the fact that $\xi_n(\bk)=\xi_n(-\bk)$ that the expectation value of any TR-odd observable vanishes. In particular, the average spin density yields $\langle s_{\mu}\rangle=0$. It is easy to see that the spin torque, although TR-even, also averages to zero in the bulk:
\begin{eqnarray*}
    \langle \tau_{\mu}\rangle &=& \frac{i}{\hbar}\sum_{n}\int\frac{d^2\bk}{(2\pi)^2}\,\langle\bk,n|\bigl[\hcH(\bk),\hat{\cal S}_\mu(\bk)\bigr]|\bk,n\rangle f(\xi_n) \\
    &=& 0,
\end{eqnarray*}
as was pointed out in Sec. \ref{sec: observables-band-rep}.
In contrast, the average spin current does not necessarily vanish in a uniform equilibrium state, as we discuss in the next section for an explicit example.

%==================================================================
%==================================================================

\section{Application to $\mathbb{G}=\mathbf{C}_{4v}$}
\label{sec: applications-C4v}

To illustrate the general points of the previous section, we consider a 2D noncentrosymmetric crystal~\cite{2D-NCS-meaning} described by the point group $\mathbb{G}=\mathbf{C}_{4v}$.
Examples of materials of this symmetry include conducting interfaces or surfaces of insulating oxides~\cite{oxide-interfaces-1,oxide-interfaces-2}, ultra-thin films of chalcogenide semiconductors~\cite{2D-chalco-1,2D-chalco-2}, or metals on various substrates~\cite{metal-films-1,metal-films-2,metal-films-3}.
An additional advantage of 2D systems is that the SOC strength can be controlled by an external electric field~\cite{SOC-2D-control-1,SOC-2D-control-2}.
We also note that the point group $\mathbf{C}_{4v}$ describes the symmetry of numerous bulk noncentrosymmetric superconductors, which have been actively studied in the last two decades~\cite{NCSC-book}. 

The group $\mathbf{C}_{4v}$ is generated by a fourfold rotation $C_{4z}$ and a mirror reflection $\sigma_{y}$. 
It has five single-valued irreducible representations (irreps): the one-dimensional (1D) $\Gamma_1$ (or $A_1$), $\Gamma_2$ (or $A_2$), $\Gamma_3$ (or $B_1$), $\Gamma_4$ (or $B_2$), and the two-dimensional (2D) $\Gamma_5$ (or $E$)~\cite{Lax-book}. 
For concreteness, we focus on just two orbital manifolds: a lower one corresponding to $\Gamma_1$, which we label by $\ell=1$, and an upper one corresponding to $\Gamma_5$, which we label by $\ell=2$. 
We assume all other states to be remote in energy and, therefore, neglect them. Such an orbital structure, with the $\Gamma_1$ ($\Gamma_5$) states stemming from the atomic $p_z$ ($p_{x,y}$) orbitals, is applicable, for instance, to 2D chalcogenides~\cite{2D-chalco-1,2D-chalco-2}.
Further, the discussion can be straightforwardly extended to other combinations of one 1D and one 2D orbital manifold in a square lattice, such as a $\Gamma_4$ ($d_{xy}$) state coupled with $\Gamma_5$ ($d_{xz,yz}$) states, which describes various oxide interfaces~\cite{oxide-interfaces-1,oxide-interfaces-2,d-orbitals-1,d-orbitals-2}. 
However, note that we assume that the atomic orbitals are strongly affected by the noncentrosymmetric crystalline environment, so that the classification of the orbital states 
into $s$-wave, $p$-wave, \textit{etc}, breaks down. For example, both $p_z$- and $s$-wave states are invariant under all elements of $\mathbf{C}_{4v}$ and correspond to the $\Gamma_1$ irrep. 
The actual $\Gamma_1$ orbital states are therefore given by superpositions of $p_z$- and $s$-wave orbitals, which do not have a definite parity.

\begin{figure}
\includegraphics{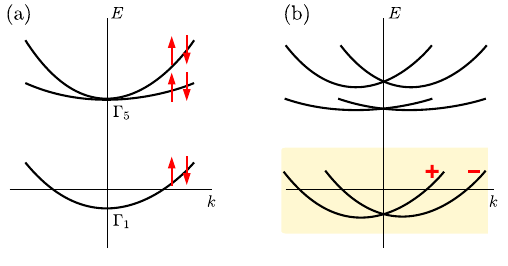}
\caption{Schematic evolution of the band structure for coupled $\Gamma_1$ and $\Gamma_5$ orbitals. (a) Spin-degenerate bands in the absence of SOC. 
(b) Effective description for $\Gamma_1$ orbitals. When diagonalizing the band structure fully, the band degeneracy is lifted by the SOC almost everywhere. In particular, the lower band is split into two helicity bands, see Eq.~(\ref{effective-Rashba-bands}).}
\label{fig: G1G5 bands}
\end{figure}

\subsection{Derivation of the effective Hamiltonian}
\label{sec: H-eff-C_4v}

According to Appendix \ref{app: p-p-example}, the microscopic Hamiltonian truncated to the two relevant orbital states is a $6\times 6$ matrix which has the form
\begin{eqnarray}
\label{H-p-p}
	&& \hat{\msH}(\bk) \nonumber\\
    && = \begin{pmatrix}
	\varepsilon_1(\bk)\hat\sigma_0 & -i\tilde ak_x\hat\sigma_0-i\tilde b\hat\sigma_y & -i\tilde ak_y\hat\sigma_0+i\tilde b\hat\sigma_x \\
	  i\tilde ak_x\hat\sigma_0+i\tilde b\hat\sigma_y & \varepsilon_2(\bk)\hat\sigma_0 & -ib\hat\sigma_z \\
	  i\tilde ak_y\hat\sigma_0-i\tilde b\hat\sigma_x & ib\hat\sigma_z & \varepsilon_2(\bk)\hat\sigma_0
	  \end{pmatrix},\nonumber\\
    &&
\end{eqnarray}
where $\varepsilon_\ell(\bk)=\epsilon_\ell+\hbar^2k^2/2m$, $m$ is the electron mass, and $k=|\bk| = \sqrt{k_x^2 +k_y^2}$. The $\Gamma$-point energy splitting due to the crystal field is given by 
$$
    \Eg=\epsilon_2-\epsilon_1>0.
$$
The real constants $\tilde a$, $b$, and $\tilde b$ characterize the usual $\bk\cdot\bp$ perturbation, the local atomic-like SOC in the 2D orbital, and the local SOC between the orbitals, respectively \cite{2D-orbital-isotropy}. Moreover, the above Hamiltonian does not include the intrinsic Rashba SOC terms in $\varepsilon_\ell(\bk)$, which are usually small~\cite{Winkler-book,kp-book,Kane56,Kane57} and would lead to less comprehensible expressions. 

The band structure near the $\Gamma$ point obtained from Eq.~(\ref{H-p-p}) is sketched in Fig. \ref{fig: G1G5 bands}. In the absence of SOC, the Hamiltonian takes the form
\begin{equation}
\label{H-p-p-no-SOC}
    \hat{\msH}(\bk) = \begin{pmatrix}
	\varepsilon_1(\bk) & -i\tilde ak_x & -i\tilde ak_y \\
	  i\tilde ak_x & \varepsilon_2(\bk) & 0 \\
	  i\tilde ak_y & 0 & \varepsilon_2(\bk)
	  \end{pmatrix}\otimes\hat\sigma_0,
\end{equation}
so that the bands are spin degenerate at each $\bk$. The band dispersions are given by
\begin{eqnarray}
\label{xi-123456}
    && \xi_{1,2}=\frac{\epsilon_1+\epsilon_2}{2}+\frac{\hbar^2k^2}{2m}-\sqrt{\left(\frac{\Eg}{2}\right)^2+\tilde a^2k^2}, \nonumber\\
    && \xi_{3,4}=\epsilon_2+\frac{\hbar^2k^2}{2m}, \nonumber\\
    && \xi_{5,6}=\frac{\epsilon_1+\epsilon_2}{2}+\frac{\hbar^2k^2}{2m}+\sqrt{\left(\frac{\Eg}{2}\right)^2+\tilde a^2k^2}. 
\end{eqnarray}
At the $\Gamma$ point, the upper four bands originating from the $\Gamma_5$ orbital remain degenerate. At $\bk\neq\bm{0}$, the orbital degeneracy is lifted by the $\bk\cdot\bp$ perturbations, as shown in Fig. \ref{fig: G1G5 bands}a.

When the SOC is included, i.e. $b,\tilde b\neq 0$, we expect the spin degeneracy to be lifted, except at $\bk=\bm{0}$ and other TR invariant points, where the Kramers degeneracy is still preserved. This is shown in Fig. \ref{fig: G1G5 bands}b. We do not attempt here to obtain the full band structure by direct diagonalization of the Hamiltonian in Eq.~(\ref{H-p-p}). Instead, we proceed in two steps as outlined in the previous section. First, we reduce the $6\times 6$ microscopic Hamiltonian to an effective $2\times 2$ one acting in the essential orbital subspace. Second, we diagonalize the effective Hamiltonian to find the band structure. 
A similar procedure has been used in the literature to obtain an effective Rashba SOC in various systems~\cite{Kane57,RS59,LYV96,Peter00,Bahramy11,Kim13,Shanavas14,Fu20}. We recap it here to set the stage for the derivation of the spin current in Sec. \ref{sec: spin-current-C4v-equilibrium}.

We assume that only the lower bands originating from $\Gamma_1$ cross the chemical potential and are therefore essential.
Following Sec. \ref{sec: effective H formalism}, we derive the effective Hamiltonian in the lower-band subspace by iteratively eliminating the inter-orbital couplings from Eq.~(\ref{H-p-p}). 
To second order, the solution of Eq.~(\ref{Q-1st-order-final}) yields the transformation matrix
\begin{equation}
\label{U-C4v}
	\hU(\bk) %=e^{i\hQ(\bk)}
    =\begin{pmatrix}
	\hbbone_{1}\otimes\hat\sigma_0-\dfrac{1}{2}\hq\hq^\dagger & i\hq \\
	i\hq^\dagger & \hbbone_{2}\otimes\hat\sigma_0-\dfrac{1}{2}\hq^\dagger\hq \\
	\end{pmatrix}+{\cal O}(\hq^3),
\end{equation}
where 
$$
  \hq(\bk) = -i\frac{\hat h_{12}(\bk)}{\Eg} 
      \begin{pmatrix}
          \hat\sigma_0 & i(b/\Eg)\hat\sigma_z \\
          -i(b/\Eg)\hat\sigma_z & \hat\sigma_0
      \end{pmatrix}+{\cal O}\left(\frac{b^2}{\Eg^2}\right)
$$
and $\hat h_{12}$ is the $2\times 4$ inter-orbital coupling block in Eq.~(\ref{H-p-p}).
We assume that $\Eg$ is the largest energy scale in the problem, so that in the last expression as well as in the formulas below in this section we keep only the leading order in $b/\Eg$, with the complete expressions given in Appendix \ref{app: SJT-c-4v}. 

Substituting the transformation matrix of Eq.~(\ref{U-C4v}) in Eq.~(\ref{H-eff-2nd-order-final}) and counting the energy from $\epsilon_1$, we arrive at the effective Hamiltonian in the lower orbital,
\begin{equation}
\label{H-eff-band-1}
	\hcH(\bk)\equiv\hcH_1(\bk)=\varepsilon(\bk)\hat\sigma_0+\bgam(\bk)\cdot\hbs.
\end{equation}
In the first term, the inter-orbital couplings renormalize the effective mass,
\begin{equation}
\label{varepsilon-eff-band-1}
  \varepsilon(\bk)=\frac{\hbar^2k^2}{2m^*}
\end{equation}
with
\begin{equation}
\label{eff-mass-band-1}
    \frac{1}{m^*}=\frac{1}{m}\left(1-\frac{2m\tilde a^2}{\hbar^2\Eg}\right).
\end{equation}
The second term has the same form as in the phenomenological expression (\ref{general-Rashba-model}), with 
\begin{equation}
\label{gamma-eff-band-1}
  \bgam(\bk)=\aR(k_y,-k_x,0)
\end{equation}
and
\begin{equation}
\label{alpha-R-band-1}
    \aR=\frac{2\tilde a\tilde b}{\Eg}.
\end{equation}
Equation~\eqref{gamma-eff-band-1} describes the effective Rashba SOC generated by the inter-orbital transitions to second order in $\tilde a$ and $\tilde b$. We assume $\aR>0$.
In general, the effective Hamiltonian can contain terms of higher order in $\bk$, which will appear after further iterations in the inter-orbital matrix elements~\cite{Winkler-book,kp-book}.
Other characteristics of the lower band, such as the spin current, are also modified by the coupling with the upper band, as we will discuss below. 

The Hamiltonian in Eq.~(\ref{H-eff-band-1}) can be diagonalized producing two nondegenerate helicity bands 
\begin{equation}
\label{effective-Rashba-bands}
    \xi_\pm(\bk)=\varepsilon(\bk)\pm|\bgam(\bk)|=\frac{\hbar^2k^2}{2m^*}\pm\aR k,
\end{equation}
see Fig. \ref{fig: G1G5 bands}. Introducing $k_{{\rm F},0}=\sqrt{2m^*\mu/\hbar^2}$, we find two circular Fermi surfaces with the radii given by
\begin{equation}
\label{k-pm}
  k_{{\rm F},\pm}=\sqrt{k_{{\rm F},0}^2+\left(\frac{m^*\aR}{\hbar^2}\right)^2}\mp\frac{m^*\aR}{\hbar^2}.
\end{equation}
The eigenstates corresponding to the bands with helicity $\lambda = \pm$ have the form
\begin{equation}
\label{helicity-eigenstates}
  |\bk,+\rangle=\begin{pmatrix}
                \cos\dfrac{\theta}{2} \\
                e^{i\varphi}\sin\dfrac{\theta}{2}
                \end{pmatrix},\quad
  |\bk,-\rangle=\begin{pmatrix}
                e^{-i\varphi}\sin\dfrac{\theta}{2} \\
                -\cos\dfrac{\theta}{2}
                \end{pmatrix},
\end{equation}
where the angles $\theta(\bk)$ and $\varphi(\bk)$ parameterize the SOC as
\begin{equation}
\label{gamma-parameter}
  \bgam=|\bgam|(\sin\theta\cos\varphi,\sin\theta\sin\varphi,\cos\theta).
\end{equation} 
In particular, for the SOC of the form given in Eq.~(\ref{gamma-eff-band-1}), we have $\theta=\pi/2$ and $\varphi=\arg\bk-\pi/2$.

\subsection{Spin current}
\label{sec: spin-current-C4v-equilibrium}

From Eqs.~(\ref{S-mu-exact}), (\ref{J-mu-i-exact}), and (\ref{H-p-p}), we obtain the microscopic spin-current operator truncated to the $\Gamma_1$ and $\Gamma_5$ states,
\begin{eqnarray}
\label{J-mu-i-d-d}
	&& \hat{\msJ}_{\mu,i}(\bk) = \begin{pmatrix}
	\dfrac{\hbar k_i}{m}\hat\sigma_\mu & -\dfrac{i\tilde a}{\hbar}\delta_{ix}\hat\sigma_\mu & -\dfrac{i\tilde a}{\hbar}\delta_{iy}\hat\sigma_\mu \\ 
	\dfrac{i\tilde a}{\hbar}\delta_{ix}\hat\sigma_\mu & \dfrac{\hbar k_i}{m}\hat\sigma_\mu & 0 \\
	\dfrac{i\tilde a}{\hbar}\delta_{iy}\hat\sigma_\mu & 0 & \dfrac{\hbar k_i}{m}\hat\sigma_\mu
	\end{pmatrix} \nonumber \\
    && \quad = \begin{pmatrix}
	\hat{\msJ}_{\mu,i}^{11}(\bk) & 0 \\
	0 & \hat{\msJ}_{\mu,i}^{22}(\bk)
	\end{pmatrix}
	+\begin{pmatrix}
	0 & \hat{\msJ}_{\mu,i}^{12}(\bk) \\
	\hat{\msJ}_{\mu,i}^{12,\dagger}(\bk) & 0
       \end{pmatrix}.
\end{eqnarray}
In the second line we explicitly separated the intra- and inter-orbital contributions. Applying the unitary transformation, Eq.~(\ref{U-C4v}), we find the effective spin~current operator in the lower-orbital manifold,
\begin{eqnarray}
      \hat{\cal J}_{\mu,i} \equiv (\hU^{-1}\hat{\msJ}_{\mu,i}\hU)_{11}=\hat{\msJ}_{\mu,i}^{11}+i\hat{\msJ}_{\mu,i}^{12}\hq^\dagger-i\hq\hat{\msJ}_{\mu,i}^{12,\dagger} \nonumber\\
      -\frac{1}{2}\bigl\{\hat{\msJ}_{\mu,i}^{11},\hq\hq^\dagger\bigr\}+\hq\hat{\msJ}_{\mu,i}^{22}\hq^\dagger+...\, .
\end{eqnarray}
To second order in the inter-orbital couplings and keeping only the leading terms in $b/\Eg$, we obtain 
\begin{eqnarray}
\label{J-mu-i-dd-final}
    \hat{\cal J}_{x,i}(\bk) &=& \frac{\hbar k_i}{m^*}\hat\sigma_x+\frac{\aR}{\hbar}\delta_{iy}\hat\sigma_0 \nonumber \\
	&&	-\frac{2\tilde b^2}{\Eg^2}\frac{\hbar k_i}{m}\hat\sigma_x-\frac{2b\tilde a^2}{\hbar\Eg^2}(\delta_{ix}k_y-\delta_{iy}k_x)\hat\sigma_y \nonumber \\ && +\frac{2b\aR}{\Eg^2}\frac{\hbar k_ik_y}{m}\hat\sigma_0, \nonumber \\
    \hat{\cal J}_{y,i}(\bk) &=& \frac{\hbar k_i}{m^*}\hat\sigma_y-\frac{\aR}{\hbar}\delta_{ix}\hat\sigma_0 \nonumber \\
	  &&	-\frac{2\tilde b^2}{\Eg^2}\frac{\hbar k_i}{m}\hat\sigma_y+\frac{2b\tilde a^2}{\hbar\Eg^2}(\delta_{ix}k_y-\delta_{iy}k_x)\hat\sigma_x \nonumber \\ && -\frac{2b\aR}{\Eg^2}\frac{\hbar k_ik_x}{m}\hat\sigma_0, \nonumber \\
    \hat{\cal J}_{z,i}(\bk) &=& \frac{\hbar k_i}{m^*}\hat\sigma_z \nonumber \\ 
	&&	-\frac{4\tilde b^2}{\Eg^2}\frac{\hbar k_i}{m}\hat\sigma_z,
\end{eqnarray}
see Appendix \ref{app: SJT-c-4v} for the complete expressions. Here we used Eqs.~(\ref{eff-mass-band-1}-\ref{alpha-R-band-1}) for the effective mass and the effective SOC. 
The first line in each component is the contribution which would be obtained by applying the standard definition, Eq.~(\ref{spin current-vertex-Rashba}), of the spin current to the effective Hamiltonian, Eq.~(\ref{H-eff-band-1}).  
The second and third lines give the explicit expressions for the corrections $\delta\hat{\cal J}_{\mu,i}$, see Eq.~(\ref{spin current-eff-general}). 

The spin-current operator, Eq.~(\ref{J-mu-i-dd-final}), can be represented in the form $\hat{\cal J}_{\mu,i}(\bk)=x_{\mu,i}(\bk)\hat\sigma_0+\bm{y}_{\mu,i}(\bk)\cdot\hbs$,
where $x_{\mu,i}(\bk)$ is even in $\bk$ and $\bm{y}_{\mu,i}(\bk)$ is odd in $\bk$, in agreement with the TR constraint, Eq.~(\ref{SJT-eff-TR-constraint}). From Eq.~(\ref{SJT-average}), using the eigenstates given by Eq.~(\ref{helicity-eigenstates}), we obtain the average spin current in a uniform equilibrium state,
\begin{equation}
\label{observable-P-average} 
  \langle j_{\mu,i}\rangle = \sum_{\lambda=\pm}\int\frac{d^2\bk}{(2\pi)^2}\,\left[x_{\mu,i}(\bk)+\lambda\bm{y}_{\mu,i}(\bk)\cdot\hat{\bgam}(\bk)\right]f(\xi_\lambda).
\end{equation}
Substituting here the effective SOC from Eq.~(\ref{gamma-eff-band-1}), it is easy to show that there are only two nonzero components, namely
\begin{equation}
\label{j-xy-J_s}
  \langle j_{x,y}\rangle=-\langle j_{y,x}\rangle=J_s(T),
\end{equation}
where
\begin{multline}
\label{J_s-general}
  J_s(T) = \frac{\aR}{2\pi\hbar}\int_0^\infty dk\,k\left(1+\frac{\hbar^2b}{m\Eg^2}k^2\right)\left(f_++f_-\right) \\
       +\frac{\hbar}{4\pi m^*}\left(1-\frac{2\tilde b^2}{\Eg^2}-\frac{2mb\tilde a^2}{\hbar^2\Eg^2}\right)\int_0^\infty dk\,k^2\left(f_+-f_-\right)
\end{multline}
and $f_\lambda=f(\xi_\lambda)$ is the Fermi distribution function in the $\lambda$th helicity band. It is important to remember that our expressions for the effective Hamiltonian and the spin current operator are valid to second order in the inter-orbital couplings $\tilde a$ and $\tilde b$, therefore $J_s(T)$ must also be evaluated within the same approximation.

At zero temperature, the integrals in Eq.~(\ref{J_s-general}) can be calculated analytically and we obtain
\begin{equation}
\label{J_s-final}
  J_s(T=0)=\frac{\hbar k_{F,0}^4b\aR}{4\pi m\Eg^2}=\left(\frac{\mu}{\Eg}\right)^2\frac{mb}{\pi\hbar^3}\,\aR.
\end{equation}
Note that the structure of this expression is qualitatively different from Eq.~(\ref{j-xy-Rashba}). 
In particular, the equilibrium spin current depends on the chemical potential and thus on the electron concentration. 
Also, in contrast to Eq.~(\ref{j-xy-Rashba}), the bulk spin current is \textit{linear} in the strength of the Rashba SOC. It was shown in Ref.~\onlinecite{Sabl08} that the spin current flowing near the edges of a sample calculated using the conventional definition, Eq.~(\ref{spin current-vertex-Rashba}), also depends linearly on $\aR$. Due to the rapid variation of the lattice potential near a surface, our derivation of the effective spin-current operator is not immediately applicable to such a nonuniform situation, which presents an interesting open question. 

Our result, Eq.~(\ref{J_s-final}), shows that the leading contribution to the equilibrium spin current comes from the $\delta\hcJ^\mu_i$ term in Eq.~(\ref{spin current-eff-general}). 
We also note that if we had applied the standard definition, Eq.~(\ref{spin current-vertex-Rashba}), to the effective Hamiltonian, Eq.~(\ref{H-eff-band-1}), then we would have obtained an equilibrium spin current proportional to $\alpha_{\rm R}^3$, as in Eq.~(\ref{j-xy-Rashba}). It follows from Eq.~(\ref{gamma-eff-band-1}) that such a contribution is of sixth order in the inter-orbital couplings and must therefore be set to zero within the accuracy of our calculation. 

If the coupling of the essential orbital state with \textit{all} other orbitals is completely neglected, which formally corresponds to setting $\Eg\to\infty$, then $m^*=m$ and $\aR=0$. Both the effective Hamiltonian of Eq.~(\ref{H-eff-band-1}) and the spin current operator in Eq.~(\ref{J-mu-i-dd-final}) then take the trivial form $\hcH(\bk)=\hbar^2k^2/2m$ and $\hat{\cal J}_{\mu,i}(\bk)=(\hbar k_i/m)\hat\sigma_\mu$, 
respectively, and we obtain from Eq.~(\ref{J_s-general}) that $J_s=0$, since there is no effective SOC. In the absence of inter-orbital transitions, a nonzero spin current can only come from the \textit{intrinsic} antisymmetric SOC described by the $k_y\hat\sigma_x-k_x\hat\sigma_y$ terms in the intra-orbital blocks of the general microscopic Hamiltonian, which we neglected above. For the calculation of such an intrinsic spin current, see Appendices \ref{app: trivial-topological-toy} and \ref{app: isolated orbital}.

%==================================================================
%==================================================================

\section{Conclusions}
\label{sec: Conclusions}

We have shown that the standard definition of the spin current in terms of an effective Hamiltonian, Eq.~(\ref{spin current-vertex-Rashba}), cannot in general be justified microscopically. 
Focusing on a few essential bands, one can describe electrons in a crystal lattice in terms of effective Hamiltonians with progressively smaller dimensions, which are obtained from the infinitely-dimensional exact microscopic Hamiltonian by iterative elimination of the interband couplings. The same procedure using the same approximations can be applied to other observables, producing the effective operators of spin current, spin torque, \textit{etc}, in the essential band subspace. While the intuitive definition of the spin current operator, $\hat{\msJ}_{\mu,i}=(1/2\hbar)\{\partial\hat{\msH}/\partial k_i,\hat{\sigma}_\mu\}$, is valid for the full microscopic Hamiltonian, eliminating the distant bands' contributions yields a considerably more complicated expression for the effective spin current, Eq. (\ref{spin current-eff-general}), which is not determined by the effective Hamiltonian alone.

As an illustration of these points, we used a simple model of coupled 1D and 2D orbital states in a 2D noncentrosymmetric crystal and explicitly demonstrated how the inter-orbital hybridization generates 
an effective antisymmetric SOC in the essential band as well as additional terms in the spin-current operator. We have found that it is these novel terms that actually \textit{dominate} the equilibrium spin current in a uniform system. Although we have calculated the spin current in a tetragonal crystal, it is straightforward to extend our analysis to other symmetries, e.g., to hexagonal crystals such as graphene, transition-metal dichalcogenides, and similar materials~\cite{graphene-review-2009,2D-vdW-materials-2016}.

Our main goal was to develop a conceptual framework for calculating spin currents, without focusing on any particular material or comparison with experiment. However, one can show that, for a given material, the expressions Eqs.~(\ref{j-xy-Rashba}) and (\ref{J_s-final}) predict very different magnitudes of the equilibrium spin current. Their ratio can be expressed in terms of  parameters obtained from experiment or from band structure calculations,
$$
  \frac{J^{\mathrm{Eq.(52)}}_s}{J^{\mathrm{Eq.(4)}}_s}\propto \left(\frac{\mu}{\Eg}\right)^2 \frac{\hbar^2b}{m\alpha_{\rm R}^2}\propto \left(\frac{\mu}{\Eg}\right)^2\frac{\mu b}{\ER^2}.
$$
Here $\ER=2\aR k_{{\rm F},0}$ is the band splitting due to the Rashba SOC and we have neglected the difference between the effective and bare electron masses. For a rough estimate, one can assume that the chemical potential and the band gap are of the same order of magnitude, and that the SOC energy scales, $b$ and $\ER$, are also of the same order of magnitude, both being smaller than either $\mu$ or $\Eg$. Then, we find $J^{\mathrm{Eq.(52)}}_s/J^{\mathrm{Eq.(4)}}_s\propto\mu/\ER\gg 1$, in other words our calculation predicts a much larger equilibrium spin current in the bulk than the one based on the conventional definition, Eq.~(\ref{spin current-vertex-Rashba}).

In this work we focused on the application of the modified definition of the spin current, Eq. (\ref{spin current-eff-general}), to metallic systems in equilibrium. One can expect 
that calculations of other quantities requiring an explicit form of the spin-current operator will also be affected, in particular, the spin conductivity and the spin Hall conductivity using the Kubo formula. 
Further applications include calculating the spin currents, both equilibrium and non-equilibrium, in superconductors and superfluids~\cite{SC-spintronics-1,SC-spintronics-2,HHHN19}. We leave all these questions for future work.

\begin{acknowledgments}
This work was supported by a Discovery Grant 2021-03705 from the Natural Sciences and Engineering Research Council of Canada (KS) and the Swiss National Science Foundation (SNSF) through Division II (No. 184739) (MS).  MHF acknowledges support from the Swiss National Science Foundation (SNSF) through Division II (number 207908).
KS is grateful to the Institute for Theoretical Physics, ETH Zurich for hospitality and to the Pauli Center for Theoretical Studies for financial support.
\end{acknowledgments}

%==================================================================
%==================================================================

\appendix

\section{Generalized $\bk\cdot\bp$ method}
\label{app: k-p-theory}

The Hamiltonian of non-interacting electrons in a perfectly periodic TR-invariant crystal has the following form: 
\begin{equation}
\label{H-microscopic}
    \hat H=\frac{\hbp^2}{2m}+U(\br)+\frac{\hbar}{4m^2c^2}\hbs[\bm{\nabla}U(\br)\times\hbp],
\end{equation}
where $\hbp=-i\hbar\bm{\nabla}$ is the momentum operator, $U(\br)$ is the crystal lattice potential, and the last term describes the electron-lattice SOC~\cite{no-ext-E}. 
In order to obtain the band structure and derive the spin current and the spin torque operators, we use the $\bk\cdot\bm{p}$ method modified in the presence of SOC~\cite{Winkler-book,kp-book}. 

From Eq.~(\ref{H-microscopic}) we obtain the reduced Hamiltonian at the wave vector $\bk$:
\begin{equation}
\label{H-reduced}
    \hat H_{\bk}=\hat H_0+\frac{\hbar^2k^2}{2m}+\delta\hat H_{\bk},
\end{equation}
where $\hat H_0=\hbp^2/2m+U(\br)$ and
\begin{eqnarray}
\label{delta H-k}
    \delta\hat H_{\bk}=\frac{\hbar}{m}(\bk\cdot\hat{\bp})\hat\sigma_0+\frac{\hbar}{4m^2c^2}\hbs[\bm{\nabla}U(\br)\times\hat{\bp}] \nonumber \\
    +\frac{\hbar^2}{4m^2c^2}\hbs[\bm{\nabla}U(\br)\times\bk]
\end{eqnarray}
is the perturbation which includes the usual $\bk\cdot\bm{p}$ term along with the relativistic corrections due to SOC. The eigenstates of $\hat H_{\bk}$ are spin-$1/2$ spinors having the same periodicity as the Bravais lattice of the crystal. The last term in Eq. (\ref{delta H-k}) represents a relativistically small correction to the momentum of electrons and is usually neglected~\cite{Kane56,Kane57,Winkler-book,kp-book}.
We keep this term in the general symmetry analysis below, but will drop it in most of the model calculations.

Note that $\hat H_0$ is the reduced Hamiltonian at the $\Gamma$ point which does not include the SOC.
Its eigenvalues, which are found from the equation $\hat H_0|\ell q\alpha\rangle=\epsilon_\ell|\ell q\alpha\rangle$, are at least twofold degenerate due to spin. 
The eigenstates have the form $\langle\br,\sigma|\ell q\alpha\rangle=f_{\ell q}(\br)\langle\sigma|\alpha\rangle$ and are labelled by the ``orbital'' index 
$\ell$, an additional index $q=1,...,d_\ell$ enumerating degenerate states within the $\ell$th orbital, and the spin index $\alpha=\uparrow,\downarrow$, 
with $\langle\sigma|\alpha\rangle=\delta_{\alpha\sigma}$ being the components of the basis spinors. 
The lattice-periodic functions $f_{\ell q}(\br)$, which are called the orbital states, transform according to a $d_\ell$-dimensional single-valued irrep $\gamma_\ell$ of the crystal point group $\mathbb{G}$:
\begin{equation}
\label{f-transform-g}
    g: f_{\ell q}(\br)\to f_{\ell q}(g^{-1}\br)=\sum_{q'=1}^{d_\ell}f_{\ell q'}(\br)D_{\ell,q'q}(g),
\end{equation}
where $\hat D_\ell(g)\equiv\hat D^{(\gamma_\ell)}(g)$ is the irrep matrix. In particular, for 1D orbitals we have $f_\ell(g^{-1}\br)=\chi_\ell(g)f_\ell(\br)$, where $\chi_\ell$ is the group character. 
If the point group contains inversion, then the orbital states have a definite parity: $f_{\ell q}(-\br)=p_\ell f_{\ell q}(\br)$, where $p_\ell=\pm 1$. Under time reversal, $f_{\ell q}(\br)\to f_{\ell q}^*(\br)$.

Next, we use the orbital states to introduce at each $\bk$ the complete and orthonormal set of the LK functions~\cite{LK55}:
\begin{equation}
\label{Luttinger-Kohn-basis}
  \langle\br,\sigma|\bk,L\rangle\equiv\langle\br,\sigma|\bk,\ell q\alpha\rangle=\frac{1}{\sqrt{\cal V}}e^{i\bk\br}f_{\ell q}(\br)\delta_{\alpha\sigma},
\end{equation}
where $L$ is the shorthand notation for the orbital and spin labels. Thus, the LK basis is constructed from the orbital states affected by the crystal field but not the SOC. 
Note that the LK functions are different from the Bloch eigenstates of the microscopic Hamiltonian (\ref{H-microscopic}). 
The latter are denoted by $\psi_{\bk,n}(\br,\sigma)\equiv\langle\br,\sigma|\bk,n\rangle$ and satisfy the equation $\hat H|\bk,n\rangle=\xi_n(\bk)|\bk,n\rangle$, where $\xi_n(\bk)$ are the band dispersion functions. 
The band eigenstates $|\bk,n\rangle$ are spinors whose spin-up and spin-down components are both nonzero in the presence of SOC.

The band structure at an arbitrary $\bk$ can be found by diagonalizing the following matrix:
\begin{eqnarray}
\label{H-LL}
    \msH_{LL'}(\bk) = \langle\bk,L|\hat H|\bk,L'\rangle = \frac{1}{{\cal V}}\langle\ell q\alpha|\hat H_{\bk}|\ell'q'\alpha'\rangle \nonumber\\
    = \varepsilon_\ell(\bk)\delta_{\ell\ell'}\delta_{qq'}\delta_{\alpha\alpha'}+M_{\ell\ell',qq'}(\bk)\delta_{\alpha\alpha'} \nonumber\\
    +\bm{L}_{\ell\ell',qq'}(\bk)\bm{\sigma}_{\alpha\alpha'},
\end{eqnarray}
where
\begin{equation}
\label{varepsilon-def-appendix}
	\varepsilon_\ell(\bk)=\epsilon_\ell+\frac{\hbar^2k^2}{2m},
\end{equation}
$m$ is the electron mass,
\begin{eqnarray}
\label{M-def}
  && M_{\ell\ell',qq'}(\bk) = -i\bm{A}_{\ell\ell',qq'}\cdot\bk,\\
\label{L-def}
  && \bm{L}_{\ell\ell',qq'}(\bk) = -i\bm{B}_{\ell\ell',qq'}+\bm{C}_{\ell\ell',qq'}\times\bk, 
\end{eqnarray}
and
\begin{eqnarray}
  && \bm{A}_{\ell\ell',qq'}=\frac{\hbar^2}{m}\frac{1}{\upsilon}\int d^3\br\, f^*_{\ell q}\bm{\nabla} f_{\ell'q'}, \nonumber \\
\label{ABC-def}
  && \bm{B}_{\ell\ell',qq'}=\frac{\hbar^2}{4m^2c^2}\frac{1}{\upsilon}\int d^3\br\,f^*_{\ell q}(\bm{\nabla}U\times\bm{\nabla}f_{\ell'q'}), \nonumber \\
  && \bm{C}_{\ell\ell',qq'}=\frac{\hbar^2}{4m^2c^2}\frac{1}{\upsilon}\int d^3\br\,f^*_{\ell q}(\bm{\nabla}U)f_{\ell'q'}.
\end{eqnarray}
The integrations here are performed over the crystal unit cell of volume $\upsilon$, with the orbital functions normalized as follows: 
\begin{equation}
\label{flq-normalization}
	\int d^3\br\, f^*_{\ell q}f_{\ell'q'}=\upsilon\delta_{\ell\ell'}\delta_{qq'}.
\end{equation}
All effects of SOC are contained in the matrix $\hat{\bm{L}}(\bk)$, whereas $\hat{M}(\bk)$ describes the usual $\bk\cdot\bm{p}$ perturbation. 
The $\bm{A}$ integrals come from the first term in Eq.~(\ref{delta H-k}) and are just the matrix elements of momentum. The $\bm{B}$ and $\bm{C}$ integrals come from the second and third terms in Eq.~(\ref{delta H-k}), respectively. 

The elements of the $d_\ell\times d_{\ell'}$ matrices $\hat{\bm{A}}_{\ell\ell'}$, $\hat{\bm{B}}_{\ell\ell'}$, and $\hat{\bm{C}}_{\ell\ell'}$ can be calculated using 
the microscopic expressions in Eq.~(\ref{ABC-def}) or regarded as phenomenological parameters, which satisfy a number of constraints imposed by the symmetry of the system, see an example in Sec.~\ref{app: p-p-example}.
If the orbital irreps are real, then it is easy to show that $\hat{\bm{A}}_{\ell\ell'}$, $\hat{\bm{B}}_{\ell\ell'}$ and $\hat{\bm C}_{\ell\ell'}$ are also real and satisfy the Hermiticity constraints
\begin{equation}
\label{ABC-Hermiticity-TR}
    \hat{\bm{A}}_{\ell\ell'}=-\hat{\bm{A}}_{\ell'\ell}^\top,\quad \hat{\bm{B}}_{\ell\ell'}=-\hat{\bm{B}}_{\ell'\ell}^\top,\quad \hat{\bm{C}}_{\ell\ell'}=\hat{\bm{C}}_{\ell'\ell}^\top. 
\end{equation}
Regarding the symmetry under the point-group operations $g$, we substitute the transformation property (\ref{f-transform-g}) into the expressions (\ref{ABC-def}) and obtain the constraints
\begin{eqnarray}
\label{eq:A-g}
  && \hat{\cal R}(g)\hat{\bm{A}}_{\ell\ell'} = \hat D^\dagger_{\ell}(g)\hat{\bm{A}}_{\ell\ell'}\hat D_{\ell'}(g),\\
  && \hat{\cal R}(g)\hat{\bm{C}}_{\ell\ell'} = \hat D^\dagger_{\ell}(g)\hat{\bm{C}}_{\ell\ell'}\hat D_{\ell'}(g).\label{eq:C-g}
\end{eqnarray}
Here, $\hat{\cal R}(g)=\hat R$ if $g$ is a proper rotation $R$ and $\hat{\cal R}(g)=-\hat R$ if $g$ is an improper rotation $IR$ ($\hat R$ is the rotation matrix). 
Thus we see that $\hat{\bm{A}}_{\ell\ell'}$ and $\hat{\bm{C}}_{\ell\ell'}$ transform as invariant vectors. In contrast,
\begin{equation}
\label{B-g}
  \hat R\hat{\bm{B}}_{\ell\ell'}=\hat D^\dagger_{\ell}(g)\hat{\bm{B}}_{\ell\ell'}\hat D_{\ell'}(g),
\end{equation}
for both proper and improper rotations, i.e. $\hat{\bm{B}}_{\ell\ell'}$ transforms as an invariant pseudovector.

If the point group contains the spatial inversion $I$, then the orbital states $f_{\ell q}(\br)$ have a definite parity and, putting $g=I$ in Eqs.~\eqref{eq:A-g}, (\ref{eq:C-g}), and (\ref{B-g}), we obtain
\begin{eqnarray}
\label{ABC-I}
    & \hat{\bm{A}}_{\ell\ell'}=-p_\ell p_{\ell'}\hat{\bm{A}}_{\ell\ell'},\quad \hat{\bm{B}}_{\ell\ell'}=p_\ell p_{\ell'}\hat{\bm{B}}_{\ell\ell'},\nonumber\\ 
    & \hat{\bm{C}}_{\ell\ell'}=-p_\ell p_{\ell'}\hat{\bm{C}}_{\ell\ell'}.
\end{eqnarray}
In particular, for the intra-orbital matrix elements we have $\hat{\bm{A}}_{\ell\ell}=\bm{0}$, $\hat{\bm{B}}_{\ell\ell}=-\hat{\bm{B}}_{\ell\ell}^\top$, and $\hat{\bm{C}}_{\ell\ell}=\bm{0}$. 
Therefore, a minimal model of the SOC in a centrosymmetric crystal must include either two 1D orbitals or one 2D orbital. 

In a noncentrosymmetric crystal, the orbital states do not have a definite parity and there is no constraint as in Eq.~(\ref{ABC-I}). In particular, the momentum operator can have nonzero matrix elements between
any two orbitals, i.e. $\hat{\bm{A}}_{\ell\ell'}\neq 0$. Also, $\hat{\bm{C}}_{\ell\ell}$ can be nonzero, producing a linear-in-$\bk$ SOC even for a single orbital.

\subsection{Example: 1D orbital $+$ 2D orbital}
\label{app: p-p-example}

As an example, let us find the general form of the parameters $\bm{A}$, $\bm{B}$, and $\bm{C}$ in a 2D crystal with the point group $\mathbf{C}_{4v}$. 
This group does not contain the spatial inversion and is generated by the rotation $C_{4z}$ and the mirror reflection $\sigma_{y}$. 
It has five single-valued irreps: four 1D ($\Gamma_1$, $\Gamma_2$, $\Gamma_3$, $\Gamma_4$) and one 2D ($\Gamma_5$), see Ref. \onlinecite{Lax-book}. 

We consider a $\Gamma_1$ state (labelled as $\#1$), which is hybridized with a $\Gamma_5$ state ($\#2$). This can be used to model a combination of $p_z$ and $p_{x,y}$ orbitals (or $s$ and $p_{x,y}$, or $s$ and $d_{xz,yz}$) affected by a  noncentrosymmetric crystal field. 
The group characters in the $\Gamma_1$ irrep are given by $\chi_{\Gamma_1}(C_{4z})=\chi_{\Gamma_1}(\sigma_y)=1$, while the $\Gamma_5$ irrep matrices can be chosen as
$\hat D_{\Gamma_5}(C_{4z})=-i\hat\tau_2$ and $\hat D_{\Gamma_5}(\sigma_y)=\hat\tau_3$, where $\hat{\bm{\tau}}$ are the Pauli matrices in the 2D irrep space.

For the intra-orbital parameters, we obtain from Eq.~(\ref{ABC-Hermiticity-TR}) that $\bm{A}_{11}=\bm{0}$, $\bm{B}_{11}=\bm{0}$, but $\bm{C}_{11}=\bm{c}\neq\bm{0}$. From the point-group constraints, Eqs.~(\ref{eq:A-g})-\eqref{eq:C-g}, with
\begin{equation}
    \hat{\cal R}(C_{4z})=\hat R(C_{4z})=\begin{pmatrix}
                          0 & -1 & 0 \\
                          1 & 0 & 0 \\
                          0 & 0 & 1
                 \end{pmatrix}
\end{equation}
and
\begin{equation}
    \hat{\cal R}(\sigma_y)=-\hat R(C_{2y})=\begin{pmatrix}
                          1 & 0 & 0 \\
                          0 & -1 & 0 \\
                          0 & 0 & 1
                   \end{pmatrix},
\end{equation}
it follows that $C_{4z}\bm{c}=\bm{c}$ and $C_{2y}\bm{c}=-\bm{c}$, therefore $\bm{c}=-\gamma_1\hat{\bm{z}}$.
In the 2D orbital, we can represent the intra-orbital matrices in the form $\hat{\bm{A}}_{22}=\bm{a}(i\hat\tau_2)$ and $\hat{\bm{B}}_{22}=\bm{b}(i\hat\tau_2)$, whereas $\hat{\bm{C}}_{22}=\bm{c}_0\hat\tau_0+\bm{c}_1\hat\tau_1+\bm{c}_3\hat\tau_3$, where $\bm{a},\bm{b}$, and $\bm{c}_{0,1,3}$ are real. From Eq.~(\ref{eq:A-g}) we obtain $C_{4z}\bm{a}=\bm{a}$ and $C_{2y}\bm{a}=\bm{a}$, therefore $\bm{a}=\bm{0}$. Similarly, one can show that $\bm{c}_1=\bm{c}_3=\bm{0}$, but $\bm{c}_0=-\gamma_2\hat{\bm{z}}$. 
From Eq. (\ref{B-g}) we obtain: 
$C_{4z}\bm{b}=\bm{b}$ and $C_{2y}\bm{b}=-\bm{b}$, therefore $\bm{b}=b\hat{\bm{z}}$. In the same fashion, one can also find the inter-orbital parameters. 

Collecting everything together, the part of the microscopic Hamiltonian which describes the coupled $\Gamma_1$ and $\Gamma_5$ orbital states has the following form:   
\begin{widetext}
\begin{equation}
\label{H-p-p-general}
	\hat{\msH}(\bk)=\begin{pmatrix}
	\varepsilon_1(\bk)\hat\sigma_0+\gamma_1(k_y\hat\sigma_x-k_x\hat\sigma_y) & -i\tilde ak_x\hat\sigma_0-i\tilde b\hat\sigma_y+\tilde\gamma k_y\hat\sigma_z & 
	      -i\tilde ak_y\hat\sigma_0+i\tilde b\hat\sigma_x-\tilde\gamma k_x\hat\sigma_z & \cdots \\
	  i\tilde ak_x\hat\sigma_0+i\tilde b\hat\sigma_y+\tilde\gamma k_y\hat\sigma_z & \varepsilon_2(\bk)\hat\sigma_0+\gamma_2(k_y\hat\sigma_x-k_x\hat\sigma_y) & -ib\hat\sigma_z & \cdots \\
	  i\tilde ak_y\hat\sigma_0-i\tilde b\hat\sigma_x-\tilde\gamma k_x\hat\sigma_z & ib\hat\sigma_z & \varepsilon_2(\bk)\hat\sigma_0+\gamma_2(k_y\hat\sigma_x-k_x\hat\sigma_y) & \cdots \\
	\vdots & \vdots & \vdots & \ddots
	\end{pmatrix}.
\end{equation}
Similarly, one can obtain the microscopic Hamiltonians for other combinations of a 1D state and a $\Gamma_5$ state, which only differ from Eq. (\ref{H-p-p-general}) by the inter-orbital blocks. For instance, for
$\Gamma_4+\Gamma_5$ (which is applicable to a $d_{xy}$ orbital coupled to a $d_{xz,yz}$ orbital, both affected by a noncentrosymmetric crystal field) we have
\begin{equation}
\label{H-d-d-general}
	\hat{\msH}(\bk)=\begin{pmatrix}
	\varepsilon_1(\bk)\hat\sigma_0+\gamma_1(k_y\hat\sigma_x-k_x\hat\sigma_y) & -i\tilde ak_y\hat\sigma_0-i\tilde b\hat\sigma_x+\tilde\gamma k_x\hat\sigma_z & 
	      -i\tilde ak_x\hat\sigma_0+i\tilde b\hat\sigma_y-\tilde\gamma k_y\hat\sigma_z & \cdots \\
	  i\tilde ak_y\hat\sigma_0+i\tilde b\hat\sigma_x+\tilde\gamma k_x\hat\sigma_z & \varepsilon_2(\bk)\hat\sigma_0+\gamma_2(k_y\hat\sigma_x-k_x\hat\sigma_y) & -ib\hat\sigma_z & \cdots \\
	  i\tilde ak_x\hat\sigma_0-i\tilde b\hat\sigma_y-\tilde\gamma k_y\hat\sigma_z & ib\hat\sigma_z & \varepsilon_2(\bk)\hat\sigma_0+\gamma_2(k_y\hat\sigma_x-k_x\hat\sigma_y) & \cdots \\
	\vdots & \vdots & \vdots & \ddots
	\end{pmatrix}.
\end{equation}
\end{widetext}
The dots in Eqs. (\ref{H-p-p-general}) and (\ref{H-d-d-general}) denote the contributions from all other orbital states, which are assumed to be well separated in energy.

The $\tilde a$ terms in Eqs. (\ref{H-p-p-general}) and (\ref{H-d-d-general}) correspond to the usual $\bk\cdot\bp$ perturbation and can be nonzero for any pair of the orbital states, 
because the latter do not have a definite parity. 
The constants $b$ and $\tilde b$ characterize the local intra-orbital SOC in the 2D orbital and the local inter-orbital SOC, respectively. These leading relativistic corrections come from the first term in $\hat{\bm L}(\bk)$, see Eq.~(\ref{L-def}), which is dominated by the atomic core regions, where the gradients of the crystal potential and of the wave functions are the greatest. 
Finally, the $\gamma_{1,2}$ and $\tilde\gamma$ terms describe the intrinsic antisymmetric SOC originating from the second term in $\hat{\bm L}$. These terms can be neglected, at least in the vicinity of the $\Gamma$ point. Indeed, if the integrals in Eq.~(\ref{ABC-def}) do not vanish by symmetry, then the expressions (\ref{M-def}) and (\ref{L-def}) can be estimated as follows:
$$
  Ak\sim \frac{\hbar^2}{ma^2}\,ka,\quad B\sim E_{\rm SO},\quad Ck\sim E_{\rm SO}\,ka,\\
$$
where $a$ is the lattice spacing and $E_{\rm SO}$ is the energy scale of the atomic SOC. Therefore, the second term in $\hat{\bm L}$ is much smaller than the first one.
Setting $\gamma_{1,2}=\tilde\gamma=0$ in Eq.~(\ref{H-p-p-general}), we arrive at the model (\ref{H-p-p}). 

%==================================================================
%==================================================================

\section{Microscopic expressions for observables}
\label{app: spin current-LK}

In this appendix, we derive the expressions for the spin-current and spin-torque operators in the second-quantization representation, as well as the corresponding matrices in the LK basis. 
We start with the Schr\"odinger equation for the electron wave functions,
\begin{equation}
\label{SE-microscopic}
  i\hbar\frac{\partial}{\partial t}\begin{pmatrix}
                                          \psi_\uparrow \\
                                          \psi_\downarrow
                                   \end{pmatrix}
                             =\hat H\begin{pmatrix}
                                          \psi_\uparrow \\
                                          \psi_\downarrow
                                    \end{pmatrix},
\end{equation}
where the Hamiltonian is given by Eq.~(\ref{H-microscopic}). We introduce the particle and spin densities in the standard form,
\begin{align}
  & \rho(\br,t) = \psi_\sigma^*(\br,t)\psi_\sigma(\br,t),\nonumber \\ 
  & s_\mu(\br,t) = \psi_\sigma^*(\br,t)\sigma_{\mu,\sigma\sigma'}\psi_{\sigma'}(\br,t). \nonumber
\end{align}
Here and below, the summation over the spin indices $\sigma,\sigma'=\uparrow,\downarrow$ is assumed. Note that the spin angular momentum density is equal to $(\hbar/2)s_\mu(\br,t)$. 

A straightforward calculation using Eq.~(\ref{SE-microscopic}) shows that the continuity equation for the spin density has the form
\begin{equation}
\label{spin-density-eq}
  \frac{\partial s_\mu}{\partial t}=-\nabla_i j_{\mu,i}+\tau_\mu,
\end{equation}
where 
\begin{eqnarray}
\label{spin current-1st-quantization}
  j_{\mu,i}(\br,t)=\frac{i\hbar}{2m}\sigma_{\mu,\sigma\sigma'}[(\nabla_i\psi_\sigma^*)\psi_{\sigma'}-\psi_\sigma^*(\nabla_i\psi_{\sigma'})]\nonumber \\
  -\frac{\hbar}{4m^2c^2}e_{\mu ij}(\nabla_jU)(\psi_\sigma^*\psi_\sigma)
\end{eqnarray}
can be interpreted as the current density of the $\mu$th spin component in the $i$th spatial direction, while the source term
\begin{eqnarray}
\label{spin-torque-1st-quantization}
  \tau_\mu(\br,t) &=& \frac{i\hbar}{4m^2c^2}\sigma_{i,\sigma\sigma'} \nonumber\\
  && \times \bigl\{(\nabla_iU)[(\nabla_\mu\psi_\sigma^*)\psi_{\sigma'}-\psi_\sigma^*(\nabla_\mu\psi_{\sigma'})] \nonumber\\
  && -(\nabla_\mu U)[(\nabla_i\psi_\sigma^*)\psi_{\sigma'}-\psi_\sigma^*(\nabla_i\psi_{\sigma'})]\bigr\}
\end{eqnarray}
is identified with the spin-torque density. 
Our definition (\ref{spin current-1st-quantization}) agrees with the one obtained by interpreting the SOC term in Eq.~(\ref{H-microscopic}) 
as an SU(2) vector potential and using the invariance of $\hat H$ with respect to local non-Abelian gauge transformations, see Refs. \onlinecite{Frohlich91,Mineev92,Leurs08,Tokatly08}.

Switching to the second-quantization representation, we obtain from Eqs.~(\ref{spin current-1st-quantization}) and (\ref{spin-torque-1st-quantization}) the following expressions for
the spin-current operator:
\begin{eqnarray}
\label{spin current-operator}
	\hat j_{\mu,i}(\br)=\frac{i\hbar}{2m}\tr[(\nabla_i\hat\Psi^\dagger)\hat\sigma_\mu\hat\Psi-\hat\Psi^\dagger\hat\sigma_\mu(\nabla_i\hat\Psi)] \nonumber\\
    -\frac{\hbar}{4m^2c^2}e_{\mu ij}(\nabla_jU)\tr(\hat\Psi^\dagger\hat\Psi),
\end{eqnarray}
the spin-torque operator:
\begin{eqnarray}
\label{spin-torque-operator}
  && \hat\tau_\mu(\br) = \frac{i\hbar}{4m^2c^2} \nonumber\\  
    && \qquad \times \bigl\{(\nabla_iU)\tr[(\nabla_\mu\hat\Psi^\dagger)\hat\sigma_i\hat\Psi-\hat\Psi^\dagger\hat\sigma_i(\nabla_\mu\hat\Psi)] \nonumber\\
    && \qquad -(\nabla_\mu U)\tr[(\nabla_i\hat\Psi^\dagger)\hat\sigma_i\hat\Psi-\hat\Psi^\dagger\hat\sigma_i(\nabla_i\hat\Psi)]\bigr\},
\end{eqnarray}
and also the spin~density operator:
\begin{equation}
\label{spin-density-operator}
	\hat s_\mu(\br)=\tr(\hat\Psi^\dagger\hat\sigma_\mu\hat\Psi).
\end{equation}
Here, $\hat\Psi^\dagger(\br,\sigma)$ and $\hat\Psi(\br,\sigma)$ are the electron creation and annihilation operators, and the trace is taken in the spin space.

\subsection{Spin current}
\label{sec: spin current-LK}

From Eq.~(\ref{spin current-operator}), we obtain the average spin-current density in thermodynamic equilibrium:
\begin{eqnarray}
\label{j-mu-i-G-r-r}
  \langle j_{\mu,i}(\br)\rangle=-\frac{i\hbar}{2m}\tr[\hat\sigma_\mu(\nabla_i-\nabla'_i)\hat G(\br,\br';0^-)]_{\br'=\br} \nonumber\\
  -\frac{\hbar}{4m^2c^2}e_{\mu ij}(\nabla_jU)\tr[\hat G(\br,\br;0^-)],\quad
\end{eqnarray}
where $G_{\sigma_1\sigma_2}(\br_1,\br_2;\tau)=-\langle T_\tau\hat\Psi(\br_1,\sigma_1;\tau)\hat\Psi^\dagger(\br_2,\sigma_2;0)\rangle$ is the Matsubara Green's function of electrons in the coordinate-spin representation.
To transform into the band representation, we expand the field operators in terms of the exact band states incorporating SOC:
\begin{align}
\label{Psi-c}
    \hat\Psi(\br,\sigma) &= \sum_{\bk,n}\psi_{\bk,n}(\br,\sigma)\hat c_{\bk,n} \nonumber \\
    &= \frac{1}{\sqrt{\cal V}}\sum_{\bk,n,L}e^{i\bk\br}f_{\ell q}(\br)\delta_{\alpha\sigma}\msW_{L,n}(\bk)\hat c_{\bk,n}\;,
\end{align}
where $\psi_{\bk,n}(\br,\sigma)\equiv\langle\br,\sigma|\bk,n\rangle$ are the Bloch eigenstates of the Hamiltonian (\ref{H-microscopic}).
In the second line, the latter are further expanded in terms of the LK basis (\ref{Luttinger-Kohn-basis}) as
$|\bk,n\rangle=\sum_L|\bk,L\rangle\,\msW_{L,n}(\bk)$,
where $\hat{\msW}(\bk)$ is the unitary matrix of the orbital-to-band transformation, see Eq. (\ref{WHW}). If one takes into account $N$ orbital states, then $\hat{\msW}(\bk)$ is a $2N\times 2N$ matrix.

Substituting the expansion Eq.~(\ref{Psi-c}) into Eq.~(\ref{j-mu-i-G-r-r}), we obtain
\begin{eqnarray}
\label{j-mu-i-G-1}
  && \langle j_{\mu,i}(\br)\rangle = \frac{1}{\cal V}\sum_{\bk_{1,2},n_{1,2}} e^{i(\bk_1-\bk_2)\br}G_{n_1n_2}(\bk_1,\bk_2;0^-) \nonumber\\
  && \times \sum_{LL'}\msW^\dagger_{n_2,L}(\bk_2) \tilde{\msJ}_{\mu,i;LL'}\left(\frac{\bk_1+\bk_2}{2},\br\right) \msW_{L',n_1}(\bk_1),\qquad
\end{eqnarray}
where $G_{n_1n_2}(\bk_1,\bk_2;\tau)=-\langle T_\tau \hat c_{\bk_1n_1}(\tau)\hat c^\dagger_{\bk_2n_2}(0)\rangle$ is the Green's function in the band representation and
\begin{eqnarray*}
    && \tilde{\msJ}_{\mu,i;LL'}(\bk,\br) = \frac{\hbar k_i}{m}f^*_{\ell q}(\br)f_{\ell'q'}(\br)\sigma_{\mu,\alpha\alpha'}\\
    &&\qquad -\frac{i\hbar}{2m}[f^*_{\ell q}(\br)\nabla_if_{\ell'q'}(\br)-\nabla_if^*_{\ell q}(\br)f_{\ell'q'}(\br)]\sigma_{\mu,\alpha\alpha'}\\
    &&\qquad -\frac{\hbar}{4m^2c^2}e_{\mu ij}(\nabla_jU)f^*_{\ell q}(\br)f_{\ell'q'}(\br)\delta_{\alpha\alpha'}.
\end{eqnarray*}
Note that the expression (\ref{j-mu-i-G-1}) is formally exact if one uses the complete LK basis (\ref{Luttinger-Kohn-basis}), i.e. includes all orbital states at the $\Gamma$ point.
 
Next, we assume that the external fields affecting the system vary slowly on the scale of the lattice constant, so that the Green's function in Eq.~(\ref{j-mu-i-G-1}) 
is nonzero only if $\bk_1$ is close to $\bk_2$. Then, the exponential $e^{i(\bk_1-\bk_2)\br}$ also varies slowly and one can replace the lattice-periodic matrix $\hat{\tilde{\msJ}}_{\mu,i}$ by its spatial average:
$$
  \tilde{\msJ}_{\mu,i;LL'}(\bk,\br)\to \msJ_{\mu,i;LL'}(\bk)=\frac{1}{\upsilon}\int d^3\br\,\tilde{\msJ}_{\mu,i;LL'}(\bk,\br),
$$
where the integration is performed over the unit cell of volume $\upsilon$. Using Eq.~(\ref{ABC-def}), we obtain the spin-current matrix in the LK basis
\begin{eqnarray}
\label{bar-I}
  && \msJ_{\mu,i;LL'}(\bk) = \frac{\hbar k_i}{m}\delta_{\ell\ell'}\delta_{qq'}\sigma_{\mu,\alpha\alpha'} \nonumber\\
  && \qquad -\frac{i}{\hbar}A_{i,\ell\ell',qq'}\sigma_{\mu,\alpha\alpha'}-\frac{1}{\hbar}e_{\mu ij}C_{j,\ell\ell',qq'}\delta_{\alpha\alpha'}.
\end{eqnarray}
Comparing Eqs.~(\ref{bar-I}) and (\ref{H-LL}), we arrive at the expression (\ref{J-mu-i-exact}). 

Note that Eq.~(\ref{bar-I}) can also be written in the following form:
\begin{equation}
\label{bar-I-H-anticommutator}
  \msJ_{\mu,i;LL'}(\bk)=\frac{1}{2}\langle\bk,L|\{\hat V_i,\hat\sigma_\mu\}|\bk,L'\rangle,
\end{equation}
where
\begin{equation}
\label{velocity-operator-SOC}
  \hat{\bm{V}}=\frac{\hat\bp}{m}\hat\sigma_0+\frac{\hbar}{4m^2c^2}(\hbs\times\bm\nabla U)
\end{equation}
is the velocity operator in the presence of SOC. Although the matrix (\ref{bar-I-H-anticommutator}) has the form similar to Eq.~(\ref{spin current-vertex-Rashba}), the crucial difference is that
the former expression is exact in the complete microscopic basis, whereas the latter is only an approximation that acts in the much smaller subspace corresponding to the essential orbital states.  

\begin{widetext}
Finally, the average spin current density in equilibrium, see Eq.~(\ref{j-mu-i-G-1}), becomes
\begin{equation}
\label{j-mu-i-general} 
  \langle j_{\mu,i}(\br)\rangle=T\sum_m\frac{1}{\cal V}\sum_{\bk,\bq}e^{i\bq\br}e^{i\omega_m 0^+} \Tr\left[\hat{\msW}^{-1}\left(\bk-\frac{\bq}{2}\right)\hat{\msJ}_{\mu,i}(\bk)
	\hat{\msW}\left(\bk+\frac{\bq}{2}\right)\hat{G}\left(\bk+\frac{\bq}{2},\bk-\frac{\bq}{2};\omega_m\right)\right],
\end{equation}
\end{widetext}
where $\omega_m=(2m+1)\pi T$ is the fermionic Matsubara frequency and ``$\Tr$'' stands for the $2N$-dimensional matrix trace in the band space. 
In a spatially uniform system, we have $\hat G(\bk,\bk';\omega_m)=\delta_{\bk,\bk'}[i\omega_m-\hH_{band}(\bk)+\mu]^{-1}$, see Eq.~(\ref{WHW}). Then, summing over the Matsubara frequencies and taking the thermodynamic limit ${\cal V}\to\infty$, we obtain
\begin{equation}
\label{j-mu-i-average-general} 
  \langle j_{\mu,i}\rangle=T\sum_m\int\frac{d^3\bk}{(2\pi)^3}\,e^{i\omega_m 0^+}\Tr\left[\hat{\msJ}_{\mu,i}(\bk)\hat{\msG}(\bk,\omega_m)\right],
\end{equation}
where $\hat{\msG}(\bk,\omega_m)=[i\omega_m-\hat{\msH}(\bk)+\mu]^{-1}$ is the Green's function in the LK basis, with $\hat{\msH}(\bk)$ and $\hat{\msJ}_{\mu,i}(\bk)$ given by Eqs. (\ref{H-LL}) and (\ref{J-mu-i-exact}), respectively.

\subsection{Spin torque}
\label{sec: spin torque-LK}

From Eq.~(\ref{spin-torque-operator}), the average spin torque is given by
\begin{eqnarray}
\label{tau-mu-GF}
    && \langle\tau_\mu(\br)\rangle = \nonumber\\
    && \quad -\frac{i\hbar}{4m^2c^2}\bigl\{(\nabla_iU)\tr[\hat\sigma_i(\nabla_\mu-\nabla'_\mu)\hat G(\br,\br';0^-)] \nonumber\\
    && \quad -(\nabla_\mu U)\tr[\hat\sigma_i(\nabla_i-\nabla'_i)\hat G(\br,\br';0^-)]\bigr\}_{\br'=\br}.
\end{eqnarray}
Using the expansion (\ref{Psi-c}), we obtain
\begin{eqnarray}
\label{tau-mu-GF-1}
  && \langle\tau_\mu(\br)\rangle = \frac{1}{\cal V} \sum_{\bk_{1,2},n_{1,2}} e^{i(\bk_1-\bk_2)\br} G_{n_1n_2}(\bk_1,\bk_2;0^-) \nonumber\\
  && \times \sum_{LL'} \msW^\dagger_{n_2,L}(\bk_2) \tilde{\msT}_{\mu;LL'}\left(\frac{\bk_1+\bk_2}{2},\br\right) \msW_{L',n_1}(\bk_1),\qquad
\end{eqnarray}
where 
\begin{eqnarray*}
    && \tilde{\msT}_{\mu;LL'}(\bk,\br) \\
    && = \frac{\hbar}{2m^2c^2}[k_\mu(\nabla_iU)-k_i(\nabla_\mu U)]f^*_{\ell q}(\br)f_{\ell'q'}(\br)\sigma_{i,\alpha\alpha'}\\
    && -\frac{i\hbar}{4m^2c^2}\bigl\{(\nabla_iU)[f^*_{\ell q}(\br)\nabla_\mu f_{\ell'q'}(\br)-\nabla_\mu f^*_{\ell q}(\br)f_{\ell'q'}(\br)]\\
    && -(\nabla_\mu U)[f^*_{\ell q}(\br)\nabla_i f_{\ell'q'}(\br)-\nabla_i f^*_{\ell q}(\br)f_{\ell'q'}(\br)]\bigr\}\sigma_{i,\alpha\alpha'}.
\end{eqnarray*}
Assuming that the external fields vary slowly on the scale of the lattice constant, one can replace the lattice-periodic matrix $\hat{\tilde\msT}_\mu$ by its spatial average:
$$
  \tilde{\msT}_{\mu;LL'}(\bk,\br)\to \msT_{\mu;LL'}(\bk)=\frac{1}{\upsilon}\int d^3\br\,\tilde{\msT}_{\mu;LL'}(\bk,\br). 
$$
It follows from Eqs.~(\ref{M-def}), (\ref{L-def}), and (\ref{ABC-def}) that 
\begin{align}
\label{T-BC}
  \msT_{\mu;LL'}(\bk)&=\frac{2}{\hbar}[(-i\bm{B}_{\ell\ell',qq'}+\bm{C}_{\ell\ell',qq'}\times\bk)\times\bm{\sigma}_{\alpha\alpha'}]_\mu\nonumber\\
  &=\frac{2}{\hbar}[\bm{L}_{\ell\ell',qq'}(\bk)\times\bm{\sigma}_{\alpha\alpha'}]_\mu,
\end{align}
from which we obtain Eq.~(\ref{T-mu-exact}). The average spin-torque density in a uniform equilibrium state is given by an expression similar to Eq.~(\ref{j-mu-i-average-general}), with $\hat{\msJ}_{\mu,i}(\bk)$ replaced by $\hat{\msT}_{\mu}(\bk)$.

\subsection{Spin density}
\label{sec: spin density-LK}

From Eq.~(\ref{spin-density-operator}), the average spin density is given by
\begin{equation}
\label{s-mu-G-r-r}
  \langle s_\mu(\br)\rangle=\tr[\hat\sigma_\mu\hat G(\br,\br;0^-)],
\end{equation}
Substituting here Eq.~(\ref{Psi-c}), we obtain
\begin{eqnarray}
\label{s-mu-GF-1}
  && \langle s_\mu(\br)\rangle = \frac{1}{\cal V}\sum_{\bk_{1,2},n_{1,2}}e^{i(\bk_1-\bk_2)\br}G_{n_1n_2}(\bk_1,\bk_2;0^-) \nonumber\\
  && \hfill \times \sum_{LL'} \msW^\dagger_{n_2,L}(\bk_2)\tilde{\msS}_{\mu;LL'}\left(\br\right)\msW_{L',n_1}(\bk_1),
\end{eqnarray}
where $\tilde{\msS}_{\mu;LL'}(\br)=f^*_{\ell q}(\br)f_{\ell'q'}(\br)\sigma_{\mu,\alpha\alpha'}$.
Assuming that the external fields vary slowly on the scale of the lattice constant, one can replace the lattice-periodic matrix $\hat{\tilde\msS}_\mu$ by its spatial average:
$$
  \tilde{\msS}_{\mu;LL'}(\br)\to \msS_{\mu;LL'}=\frac{1}{\upsilon}\int d^3\br\,\tilde{\msS}_{\mu;LL'}(\br). 
$$
Using the normalization condition (\ref{flq-normalization}), we arrive at Eq.~(\ref{S-mu-exact}). The average spin density in a uniform equilibrium state is given by an expression similar to Eq.~(\ref{j-mu-i-average-general}), 
with $\hat{\msJ}_{\mu,i}(\bk)$ replaced by $\hat{\msS}_{\mu}(\bk)$.

\subsection{Charge current}
\label{sec: charge current-LK}

For reference, in this subsection we derive the representation of the charge-current operator for electrons in the LK basis. Applying the standard quantum-mechanical procedure to the Hamiltonian (\ref{H-microscopic}), we obtain the second-quantized particle-current operator:
\begin{eqnarray}
\label{charge current-operator}
	\hat j_{i}(\br) &=& \frac{i\hbar}{2m}\tr[(\nabla_i\hat\Psi^\dagger)\hat\Psi-\hat\Psi^\dagger(\nabla_i\hat\Psi)] \nonumber\\
    && -\frac{\hbar}{4m^2c^2}e_{ij\mu}(\nabla_jU)\tr(\hat\Psi^\dagger\hat\sigma_\mu\hat\Psi).
\end{eqnarray}
The charge current operator is then given by $\hat j^{\text{c}}_{i}(\br)=-e\hat j_{i}(\br)$. 

Repeating the steps from Sec.~\ref{app: spin current-LK}, the average current density in equilibrium is given by
\begin{eqnarray}
\label{charge-j-i-G-1}
  && \langle j_{i}(\br)\rangle = \frac{1}{\cal V}\sum_{\bk_{1,2},n_{1,2}} e^{i(\bk_1-\bk_2)\br}G_{n_1n_2}(\bk_1,\bk_2;0^-) \nonumber\\
  && \times \sum_{LL'} \msW^\dagger_{n_2,L}(\bk_2)\tilde{\msJ}_{i;LL'}\left(\frac{\bk_1+\bk_2}{2},\br\right) \msW_{L',n_1}(\bk_1),\qquad 
\end{eqnarray}
where
\begin{eqnarray*}
    && \tilde{\msJ}_{i;LL'}(\bk,\br) = \frac{\hbar k_i}{m}f^*_{\ell q}(\br)f_{\ell'q'}(\br)\delta_{\alpha\alpha'} \\
    && -\frac{i\hbar}{2m}[f^*_{\ell q}(\br)\nabla_if_{\ell'q'}(\br)-\nabla_if^*_{\ell q}(\br)f_{\ell'q'}(\br)]\delta_{\alpha\alpha'} \\
    && -\frac{\hbar}{4m^2c^2}e_{ij\mu}(\nabla_jU)f^*_{\ell q}(\br)f_{\ell'q'}(\br)\sigma_{\mu,\alpha\alpha'}.
\end{eqnarray*}
Replacing the last expression by its spatial average,
$$
  \tilde{\msJ}_{i;LL'}(\bk,\br)\to \msJ_{i;LL'}(\bk)=\frac{1}{\upsilon}\int d^3\br\,\tilde{\msJ}_{i;LL'}(\bk,\br), 
$$
and using Eq.~(\ref{H-LL}) we obtain
\begin{equation}
\label{particle-J-LL}
    \hat{\msJ}_i(\bk)=\frac{1}{\hbar}\frac{\partial\hat\msH(\bk)}{\partial k_i}. 
\end{equation}
It is easy to check that this matrix can also be written in the form
$$
  \msJ_{i;LL'}(\bk)=\langle\bk,L|\hat V_i|\bk,L'\rangle,
$$
where $\hat{\bm{V}}$ is the velocity operator (\ref{velocity-operator-SOC}) and $|\bk,L\rangle$ are the LK states (\ref{Luttinger-Kohn-basis}). Multiplying Eq.~(\ref{particle-J-LL}) by the electron charge $-e$, we arrive at the expression of Eq.~(\ref{charge-J-exact}).

%==================================================================
%==================================================================

\section{Unraveling the spin-current contributions in a toy model}
\label{app: trivial-topological-toy}

The simplest model of the band structure is obtained by keeping just one 1D orbital and assuming that all inter-orbital transitions can be neglected. This is equivalent to truncating the Hamiltonians (\ref{H-p-p-general}) or (\ref{H-d-d-general}) to the following form:
\begin{equation}
\label{H-toy}
  \hat{\msH}(\bk)=\frac{\hbar^2k^2}{2m}\hat\sigma_0+\alpha_0(k_y\hat\sigma_x-k_x\hat\sigma_y),
\end{equation}
where $\alpha_0=\gamma_1$ is the strength of the intrinsic antisymmetric SOC and the energy is counted from $\epsilon_1$. The spin current operator is given by
\begin{equation}
\label{toy-model-spin-J}
    \msJ_{\mu,i}(\bk)=\frac{\hbar k_i}{m}\hat\sigma_\mu+\frac{\alpha_0}{\hbar}\left(\delta_{\mu x}\delta_{iy}-\delta_{\mu y}\delta_{ix}\right)\hat\sigma_0.
\end{equation}
As discussed in Appendix \ref{app: p-p-example}, this model is rather unphysical, because the intrinsic antisymmetric SOC is small. We use this model here because (i) it looks formally the same as the Rashba model, Eq.~(\ref{general-Rashba-model}), with the microscopic parameters replaced by the effective ones: $m\to m^*$ and $\alpha_0\to\aR$, and (ii) the spin current operator definition, Eq.~(\ref{J-mu-i-exact}), is not modified by the inter-orbital mixing. 

The Hamiltonian (\ref{H-toy}) is diagonalized by the matrix
\begin{equation}
\label{W-toy}
    \hat{\msW}(\bk)=\frac{1}{\sqrt{2}} \begin{pmatrix}
        1 & ie^{-i\phi_{\bk}} \\
        -ie^{i\phi_{\bk}} & 1 
        \end{pmatrix},
\end{equation}
where $\phi_{\bk}=\arg\bk$, producing the band dispersions $\xi_\lambda(\bk)=\hbar^2k^2/2m+\lambda\alpha_0 k$ ($\lambda=\pm$ and we assume that $\alpha_0>0$). Using Eq.~(\ref{W-toy}), we obtain the spin operators in the band representation:
\begin{eqnarray}
\label{S-mu-toy-band}
    && \hat S_x(\bk) = \frac{1}{k} \begin{pmatrix}
                    k_y & -k_x e^{-i\phi_{\bk}} \\
                    -k_x e^{i\phi_{\bk}} & -k_y
                    \end{pmatrix}, \nonumber \\
    && \hat S_y(\bk) = \frac{1}{k} \begin{pmatrix}
                    -k_x & -k_y e^{-i\phi_{\bk}} \\
                    -k_y e^{i\phi_{\bk}} & k_x
                    \end{pmatrix}, \nonumber \\
    && \hat S_z(\bk) = \begin{pmatrix}
                    0 & i e^{-i\phi_{\bk}} \\
                    -i e^{i\phi_{\bk}} & 0
                    \end{pmatrix}.
\end{eqnarray}
For the spin current operator (\ref{toy-model-spin-J}) we find that only the following diagonal matrix elements are nonzero:
\begin{eqnarray}
\label{toy-j-mu-i-intraband}
    && J_{x,i}^{\lambda\lambda}(\bk)=\lambda\frac{\hbar k_i}{m}\frac{k_y}{k}+\frac{\alpha_0}{\hbar}\delta_{iy}, \nonumber \\ 
    && J_{y,i}^{\lambda\lambda}(\bk)=-\lambda\frac{\hbar k_i}{m}\frac{k_x}{k}-\frac{\alpha_0}{\hbar}\delta_{ix}.  
\end{eqnarray}
Substituting these expressions in Eq.~(\ref{j-mu-i-average-bands}) and calculating the integrals at zero temperature, we obtain the equilibrium spin current
\begin{equation}
\label{toy-model-j-xy-equilibrium}
  \langle j_{x,y}\rangle=-\langle j_{y,x}\rangle=J_s(T=0)=\frac{m^{2}}{3\pi\hbar^5}\alpha_0^3.
\end{equation}
This has the same form as the phenomenological expression (\ref{j-xy-Rashba}), with $\aR$ replaced by $\alpha_0$ and $m^*$ replaced by the electron mass $m$.

Turning to the general formula, Eq.~(\ref{spin-J-intraband}), we see that the intra-band spin currents contain two distinct contributions, the ``trivial'' one, which is given by the first term, and also the ``topological'' one, which depends on the Berry connections and is given by the second term. Below we calculate these two contributions to the net equilibrium spin current (\ref{toy-model-j-xy-equilibrium}). 

Using Eq.~(\ref{S-mu-toy-band}), the trivial (``$\text{tr}$'') contributions to the spin-current operator in the $\lambda$th band are given by
\begin{eqnarray}
\label{toy-j-trivial}
    && \left. J_{x,y}^{\lambda\lambda}(\bk)\right|_{\text{tr}}=\lambda\frac{\hbar k_y}{m}\frac{k_y}{k}+\frac{\alpha_0}{\hbar} \frac{k_y^2}{k^2}, \nonumber\\
    && \left. J_{y,x}^{\lambda\lambda}(\bk)\right|_{\text{tr}}=-\lambda\frac{\hbar k_x}{m}\frac{k_x}{k}-\frac{\alpha_0}{\hbar} \frac{k_x^2}{k^2}.
\end{eqnarray}
Here, we focused only on the components that do not vanish after the momentum integration in Eq.~(\ref{j-mu-i-average-bands}).
The topological (``$\text{top}$'') contributions have the form
$$
   \left. J_{\mu,i}^{++}(\bk)\right|_{\text{top}} = \left. J_{\mu,i}^{--}(\bk)\right|_{\text{top}} = \frac{1}{\hbar}(\xi_+-\xi_-)\,\mathrm{Im}\,(\itO^i_{+-} S_\mu^{-+}). 
$$
From Eq.~(\ref{W-toy}), we find the Berry connection matrix in the helicity-band basis:
$$
    \hat{\bm{\itO}}(\bk) = \frac{1}{2}\begin{pmatrix}
                     1 & -ie^{-i\phi_{\bk}} \\
                     ie^{i\phi_{\bk}} & -1
                    \end{pmatrix} \frac{\partial\phi_{\bk}}{\partial\bk}.
$$
Using $\partial\phi_{\bk}/\partial\bk=(-k_y,k_x)/k^2$, we finally obtain
\begin{equation}
\label{toy-j-topological}
    \left. J_{x,y}^{\lambda\lambda}(\bk)\right|_{\text{top}} =\frac{\alpha_0}{\hbar} \frac{k_x^2}{k^2}, \quad
    \left. J_{y,x}^{\lambda\lambda}(\bk)\right|_{\text{top}} =- \frac{\alpha_0}{\hbar} \frac{k_y^2}{k^2}.
\end{equation}

Next, we calculate the momentum integrals in Eq.~(\ref{j-mu-i-average-bands}) and obtain the trivial and topological contributions to the equilibrium spin current at zero temperature:
\begin{equation}
\label{toy-J_s-tr}
    J^{\text{tr}}_s(T=0)=-\frac{k_{{\rm F},0}^2\alpha_0}{4\pi\hbar}-\frac{m^2\alpha_0^3}{6\pi\hbar^5}
\end{equation}
and 
\begin{equation}
\label{toy-J_s-top}
    J^{\text{top}}_s(T=0)=\frac{k_{{\rm F},0}^2\alpha_0}{4\pi\hbar}+\frac{m^2\alpha_0^3}{2\pi\hbar^5},
\end{equation}
with $k_{{\rm F},0}=\sqrt{2m\mu/\hbar^2}$. The second terms in these expressions are smaller than the first ones by a factor of $({\cal E}_{0}/\mu)^2$, where ${\cal E}_0=2\alpha_0k_{{\rm F},0}$ is the helicity band splitting, which serves as a measure of the intrinsic SOC strength. However, the dominant terms exactly cancel each other out when both contributions are added together, resulting in the expression (\ref{toy-model-j-xy-equilibrium}) for the net spin current.

%==================================================================
%==================================================================

\section{Derivation of the effective Hamiltonian}
\label{app: derivation of H-eff}

We aim to construct an effective Hamiltonian for the essential orbital state, which is obtained by removing all inter-orbital couplings from the exact microscopic Hamiltonian by a canonical transformation.
This procedure, which is generally applicable to the matrix elements connecting different groups of degenerate, or quasi-degenerate, states in an arbitrary Hamiltonian, is known in the literature as the Luttinger-Kohn~\cite{LK55} or Schrieffer-Wolff~\cite{SW66} transformation.

We begin by representing the matrix (\ref{H-LL}) in the form $\hat{\msH}(\bk)=\hat{\msH}'(\bk)+\hat{\msH}''(\bk)$, where
\begin{equation}
\label{H-p-pp-def}
  \hat{\msH}'=\begin{pmatrix}
       \hat h_1 & 0 & \cdots \\
       0 & \hat h_2 & \cdots \\
       \vdots & \vdots & \ddots
       \end{pmatrix}, \quad
       \hat{\msH}''=\begin{pmatrix}
       0 & \hat h_{12} & \cdots \\
       \hat h_{21} & 0 & \cdots \\
       \vdots & \vdots & \ddots
       \end{pmatrix}.
\end{equation}
The matrix elements corresponding to the same orbital are included in the $2d_\ell\times 2d_\ell$ ``bare'' intra-orbital Hamiltonians
\begin{multline}
\label{h-l-gen}
	\hat h_\ell(\bk)=\varepsilon_\ell(\bk)\hat{\mathbb{1}}_\ell\otimes\hat\sigma_0+(-i\hat{\bm A}_{\ell\ell}\cdot\bk)\otimes\hat\sigma_0\\
    +(-i\hat{\bm B}_{\ell\ell}+\hat{\bm C}_{\ell\ell}\times\bk)\otimes\hbs,
\end{multline}
while the matrix elements connecting different orbitals are collected into the $2d_\ell\times 2d_{\ell'}$ blocks ($\ell\neq\ell'$)
\begin{eqnarray}
\label{delta-h-ll-gen}
	&& \hat h_{\ell\ell'}(\bk)=\hat h^\dagger_{\ell'\ell}(\bk) \nonumber\\
    && =(-i\hat{\bm A}_{\ell\ell'}\cdot\bk)\otimes\hat\sigma_0+(-i\hat{\bm B}_{\ell\ell'}+\hat{\bm C}_{\ell\ell'}\times\bk)\otimes\hbs.
\end{eqnarray}
To develop a perturbative treatment of the inter-orbital couplings, we introduce a bookkeeping parameter $\zeta$, so that the Hamiltonian (\ref{H-LL}) becomes
\begin{equation}
\label{H-zeta}
    \hat{\msH}(\bk)=\hat{\msH}'(\bk)+\zeta\hat{\msH}''(\bk),\quad 0\leq\zeta\leq 1.
\end{equation}
We assume that the energy gaps between different orbital states are much larger than the matrix elements of $\hat{\msH}''$.

We shall now try to remove the inter-orbital elements from the matrix (\ref{H-zeta}) by a unitary transformation as follows:
\begin{equation}
\label{unitary-H-H-eff}
  \hcH_{\text{eff}}(\bk)=\hU^{-1}(\bk)\hat{\msH}(\bk)\hU(\bk),\quad \hU=e^{i\hQ},
\end{equation}
where $\hQ$ is a $2N\times 2N$ Hermitian matrix, which has the same block structure as $\hat{\msH}''$. Seeking it in the form of a power series: $\hQ=\sum_{p=1}^\infty\zeta^p\hQ^{(p)}$, 
the effective Hamiltonian can be represented as
\begin{equation}
\label{Lambda-expansion}
    \hcH_{\text{eff}}=\hat{\msH}'+\zeta\hat\Lambda^{(1)}+\zeta^2\hat\Lambda^{(2)}+...,
\end{equation}
where
\begin{eqnarray*}
  && \hat\Lambda^{(1)}=i[\hat{\msH}',\hQ^{(1)}]+\hat{\msH}'',\\
  && \hat\Lambda^{(2)}=i[\hat{\msH}',\hQ^{(2)}]-\frac{1}{2}[[\hat{\msH}',\hQ^{(1)}],\hQ^{(1)}]+i[\hat{\msH}'',\hQ^{(1)}],\ ...\,. 
\end{eqnarray*}
The parameter $\zeta$ can also be used to quantify the inter-orbital contributions to other observables, such as the spin current. It will be set to unity at the end of the calculations.

Since both terms in $\hat\Lambda^{(1)}$ have only inter-orbital blocks, we find $\hQ^{(1)}$ from the equation $[\hat{\msH}',\hQ^{(1)}]=i\hat{\msH}''$. Then, eliminating the inter-orbital blocks from 
$\hat\Lambda^{(2)}=i[\hat{\msH}',\hQ^{(2)}]+i[\hat{\msH}'',\hQ^{(1)}]/2$, we find $\hQ^{(2)}$, whereas the intra-orbital blocks contribute to the effective Hamiltonians in the second order in $\hat{\msH}''$. 
Proceeding in this way, one can calculate $\hQ^{(p)}$ and $\hat\Lambda^{(p)}$ by iteration to any desired order and obtain: 
\begin{equation}
\label{H-eff-general-structure}
  \hcH_{\rm eff}(\bk)=\begin{pmatrix}
       \hcH_1(\bk) & 0 & \cdots \\
       0 & \hcH_2(\bk) & \cdots \\
       \vdots & \vdots & \ddots
       \end{pmatrix},
\end{equation}
and 
\begin{equation}
  \hQ(\bk)=\begin{pmatrix}
       0 & \hQ_{12}(\bk) & \cdots \\
       \hQ_{12}^\dagger(\bk) & 0 & \cdots \\
       \vdots & \vdots & \ddots
       \end{pmatrix},    
\end{equation}
where $\hcH_\ell(\bk)$ is a $2d_{\ell}\times 2d_{\ell}$ matrix, which can be interpreted as the effective Hamiltonian in the $\ell$th orbital band. In the second order in the inter-orbital couplings, we have
\begin{equation}
\label{H-eff-2nd-order}
    \hcH_\ell=\hat h_\ell+\frac{i}{2}\sum_{\ell'\neq\ell}\bigl[\hat h_{\ell\ell'}\hQ^{(1)}_{\ell'\ell}-\hQ^{(1)}_{\ell\ell'}\hat h_{\ell'\ell}\bigr],
\end{equation}
where $\hat h_\ell$ is given by Eq. (\ref{h-l-gen}) and the $2d_{\ell}\times 2d_{\ell'}$ matrix $\hQ^{(1)}_{\ell\ell'}$ is found from the equation 
\begin{equation}
\label{Q1-equation}
	\hat h_{\ell}\hQ^{(1)}_{\ell\ell'}-\hQ^{(1)}_{\ell\ell'}\hat h_{\ell'}=i\,\hat h_{\ell\ell'}.
\end{equation}
Although the matrix elements of the microscopic Hamiltonian $\hat{\msH}(\bk)$ are either constants or linear or quadratic functions of $\bk$, see Eqs.~(\ref{h-l-gen}) and (\ref{delta-h-ll-gen}), the effective 
Hamiltonian $\hcH_{\text{eff}}(\bk)$ can contain terms of higher orders in $\bk$, which are generated by eliminating the inter-orbital couplings. For instance, the Dresselhaus SOC, which is cubic in momentum~\cite{Dressel55},
is produced in this way. Note that $\hU(\bk)$ and $\hcH_\ell(\bk)$ are analytic functions of $\bk$, by construction.

Solution of Eq. (\ref{Q1-equation}) becomes particularly simple if the splittings between different orbital bands are much larger than all intra-orbital and inter-orbital couplings. In that case, one can replace 
$\hat h_{\ell}(\bk)\to\varepsilon_\ell(\bk)\hbbone_\ell$ on the left-hand side of Eq.~(\ref{Q1-equation}) and obtain
\begin{eqnarray*}
    && \hcH_\ell(\bk)=\hat h_\ell(\bk)+\sum_{\ell'\neq\ell}\frac{\hat h_{\ell\ell'}(\bk)\hat h_{\ell'\ell}(\bk)}{\epsilon_\ell-\epsilon_{\ell'}}+...,\\ 
   && \hQ_{\ell\ell'}(\bk)=i\frac{\hat h_{\ell\ell'}(\bk)}{\epsilon_\ell-\epsilon_{\ell'}}+....
\end{eqnarray*}
In some cases, the intra-orbital Hamiltonians $\hat h_{\ell}(\bk)$ can be treated exactly and Eq.~(\ref{Q1-equation}) can be solved without making the above assumption, see Appendix~\ref{app: SJT-c-4v}.

The expectation values of observables in a uniform equilibrium state, see, e.g., Eq.~(\ref{j-mu-i-average-general}), take the following form after the transformation (\ref{unitary-H-H-eff}):
\begin{equation}
\label{j-mu-i-effective-sum-l} 
  \langle O\rangle=T\sum_m\sum_\ell\int\frac{d^2\bk}{(2\pi)^2}\,e^{i\omega_m 0^+}\Tr\left[\hcO_{\ell\ell}(\bk)\hat{\cal G}_\ell(\bk,\omega_m)\right],
\end{equation}
where 
$\hat{\cal G}_\ell(\bk,\omega_m)=[i\omega_m-\hcH_{\ell}(\bk)+\mu]^{-1}$ and $\hcO_{\ell\ell}(\bk)$ is a $2d_\ell\times 2d_\ell$ matrix representing the effective observable in the $\ell$th state, 
see Eq.~(\ref{SJT-blocks}). Focusing on just one essential orbital state, diagonalizing the effective Hamiltonian, and summing over the Matsubara frequencies, we obtain Eq.~(\ref{SJT-average}).

%==================================================================
%==================================================================

\section{TR symmetry}
\label{app: TRI}

Assuming real orbital states, TR operation acts in the LK basis (\ref{Luttinger-Kohn-basis}) as follows:
\begin{equation}
\label{f-transform-K}
  K|\bk,L\rangle=\sum_{L'}|-\bk,L'\rangle\mathsf{D}_{L'L}(K),
\end{equation}
where 
$$
	\hat{\mathsf{D}}(K)= \begin{pmatrix}
		    \hbbone_1\otimes(-i\hat\sigma_y) & 0 & \cdots \\
		    0 & \hbbone_2\otimes(-i\hat\sigma_y) & \cdots \\
		    \vdots & \vdots & \ddots
                   \end{pmatrix}=-i\hat{\msS}_y
$$
is the matrix representation of the TR operator.
Since the Hamiltonian (\ref{H-microscopic}) is TR invariant, we have
\begin{align}
  \msH_{LL'}(\bk) &= \langle\bk,L|\hat H|\bk,L'\rangle\nonumber\\ 
  &= \langle\bk,L|K^{-1}\hat HK|\bk,L'\rangle\nonumber\\
  &=\sum_{L_1,L_2}\mathsf{D}^*_{L_1L'}(K)\msH^*_{L_2L_1}(-\bk)\mathsf{D}_{L_2L}(K).
\end{align}
Here, we used Eq.~(\ref{f-transform-K}) and the antiunitarity property $\langle i|K^{-1}|j\rangle=\langle j|K|i\rangle$. Thus, the microscopic Hamiltonian matrix satisfies the following TR invariance constraint: 
\begin{equation}
\label{H-microscopic-TR-constraint}
  \hat{\mathsf{D}}^{-1}(K)\hat{\msH}(\bk)\hat{\mathsf{D}}(K)=\hat{\msH}^*(-\bk).
\end{equation}
It is easy to see that this is equivalent to the requirement that the coefficients $\bm{A}$, $\bm{B}$, and $\bm{C}$ in Eqs.~(\ref{M-def}) and (\ref{L-def}) are real.
Applying Eq.~(\ref{H-microscopic-TR-constraint}) to the operators (\ref{S-mu-exact}), (\ref{J-mu-i-exact}), and (\ref{T-mu-exact}), we obtain:
\begin{equation}
\label{SJT-TR-constraints}
  \hat{\mathsf{D}}^{-1}(K)\hat{\msO}(\bk)\hat{\mathsf{D}}(K)=\pm\hat{\msO}^*(-\bk),
\end{equation}
where the plus sign applies to the TR-even observables $\hat{\msO}=\hat{\msJ}_{\mu,i}$ or $\hat{\msT}_{\mu}$, and the minus sign to the TR-odd observable $\hat{\msO}=\hat{\msS}_{\mu}$. 

One can show that the TR constraints in the basis obtained by the transformation (\ref{LK-transform}) have exactly the same form as the constraints (\ref{H-microscopic-TR-constraint}) and (\ref{SJT-TR-constraints}) in the LK basis. 
It is sufficient to prove that the transformation matrix $\hU$ satisfies
\begin{equation}
\label{U-K-TR-constraint}
	\hat{\mathsf{D}}^{-1}(K)\hU(\bk)\hat{\mathsf{D}}(K)=\hU^*(-\bk).
\end{equation}
According to the procedure outlined in Appendix \ref{app: derivation of H-eff}, we have $\hU=\exp[i\sum_p\zeta^p\hQ^{(p)}]$, where $\hQ^{(p)}$ are calculated perturbatively in $\hat{\msH}''$. One obtains from
Eqs.~(\ref{H-zeta}) and (\ref{H-microscopic-TR-constraint}) that $\hat{\mathsf{D}}^{-1}(K)\hQ^{(p)}(\bk)\hat{\mathsf{D}}(K)=-\hQ^{(p),*}(-\bk)$, then the property (\ref{U-K-TR-constraint}) 
follows immediately. Focusing on the effective operators in one orbital band, i.e. on the diagonal blocks of Eqs.~(\ref{LK-transform}), (\ref{spin-current-LK-transformed}), (\ref{spin-torque-LK-transformed}), 
and (\ref{spin-operator-LK-transformed}), we arrive at Eqs.~(\ref{H-eff-TR-constraint}) and (\ref{SJT-eff-TR-constraint}).

%==================================================================
%==================================================================

\section{Effective observables in $\Gamma_1$}
\label{app: SJT-c-4v}

The microscopic Hamiltonian in the six-dimensional subspace of coupled $\Gamma_1$ and $\Gamma_5$ states has the form (\ref{H-p-p}), whereas for the spin operator we have
\begin{equation}
\label{S-mu-p-p}
	\hat{\msS}_{\mu}= \begin{pmatrix}
	\hat\sigma_\mu & 0 & 0 \\ 
	0 & \hat\sigma_\mu & 0\\
	0 & 0 & \hat\sigma_\mu
	\end{pmatrix}.
\end{equation}
From Eqs.~(\ref{J-mu-i-exact}) and (\ref{T-mu-exact}), we obtain the spin-current operator
\begin{equation}
\label{J-mu-i-p-p-app}
	\hat{\msJ}_{\mu,i}(\bk)= \begin{pmatrix}
	\dfrac{\hbar k_i}{m}\hat\sigma_\mu & -\dfrac{i\tilde a}{\hbar}\delta_{ix}\hat\sigma_\mu & -\dfrac{i\tilde a}{\hbar}\delta_{iy}\hat\sigma_\mu \vspace*{3pt} \\ 
	\dfrac{i\tilde a}{\hbar}\delta_{ix}\hat\sigma_\mu & \dfrac{\hbar k_i}{m}\hat\sigma_\mu & 0 \vspace*{3pt} \\
	\dfrac{i\tilde a}{\hbar}\delta_{iy}\hat\sigma_\mu & 0 & \dfrac{\hbar k_i}{m}\hat\sigma_\mu
	\end{pmatrix}
\end{equation}
and the spin torque operator
\begin{equation}
\label{T-mu-p-p}
	\hat{\msT}_\mu(\bk)=\frac{2i}{\hbar} \begin{pmatrix}
	0 & \tilde b e_{y\mu\nu}\hat\sigma_\nu & -\tilde be_{x\mu\nu}\hat\sigma_\nu \\
	-\tilde b e_{y\mu\nu}\hat\sigma_\nu & 0 & be_{z\mu\nu}\hat\sigma_\nu \\
        \tilde be_{x\mu\nu}\hat\sigma_\nu & -be_{z\mu\nu}\hat\sigma_\nu & 0
	\end{pmatrix}.
\end{equation}
The above expressions can be represented as
$$
  \hat{\msO}(\bk)=\begin{pmatrix}
       \hat{\msO}_{11}(\bk) & 0 \\
       0 & \hat{\msO}_{22}(\bk) \\
       \end{pmatrix}
       +\begin{pmatrix}
       0 & \hat{\msO}_{12}(\bk) \\
       \hat{\msO}_{12}^\dagger(\bk) & 0 
       \end{pmatrix},
$$
where $\hat{\msO}=\hat{\msS}_\mu$, $\hat{\msJ}_{\mu,i}$, or $\hat{\msT}_\mu$. The intra-orbital blocks $\hat{\msO}_{11}$ and $\hat{\msO}_{22}$ are $2\times 2$ and $4\times 4$ matrices, respectively, 
while the inter-orbital block $\hat{\msO}_{12}$ is a $2\times 4$ matrix. We decouple the orbitals by applying the unitary transformation (\ref{U-C4v}) with 
\begin{eqnarray}
\label{q-g1g5-general}
  \hq(\bk) &\equiv& \hQ_{12}(\bk) \nonumber\\ 
  &=& -\frac{i}{\Eg^2-b^2} \hat h_{12}(\bk)
      \begin{pmatrix}
          \Eg\hat\sigma_0 & ib\hat\sigma_z \\
          -ib\hat\sigma_z & \Eg\hat\sigma_0
      \end{pmatrix},
\end{eqnarray}
and obtain the effective observables in the $\Gamma_1$ band, 
\begin{eqnarray}
\label{O11-general}
    \hcO \equiv (\hU^{-1}\hat{\msO}\hU)_{11}=\hat{\msO}_{11}+i\hat{\msO}_{12}\hq^\dagger-i\hq\hat{\msO}_{12}^\dagger \nonumber\\
    -\frac{1}{2}\bigl\{\hat{\msO}_{11},\hq\hq^\dagger\bigr\}+\hq\hat{\msO}_{22}\hq^\dagger,
\end{eqnarray}
to second order in the inter-orbital couplings. 

From Eqs.~(\ref{H-eff-2nd-order-final}) and (\ref{q-g1g5-general}), the effective Hamiltonian has the form
\begin{equation}
\label{H-eff-pp-gen}
	\hcH(\bk)=\varepsilon(\bk)\hat\sigma_0+\bgam(\bk)\cdot\hbs,
\end{equation}
where
\begin{equation}
\label{varepsilon-eff-dd-gen}
    \varepsilon(\bk)=\epsilon_1-\frac{2\tilde b^2}{\Eg-b}+\frac{\hbar^2k^2}{2m^*},
\end{equation}
with the effective mass given by 
$$
\frac{1}{m^*}=\frac{1}{m}\left[1-\frac{2m\Eg\tilde a^2}{\hbar^2(\Eg^2-b^2)}\right],
$$
and
\begin{equation}
\label{gamma-eff-pp-gen}
  \bgam(\bk)=\aR(k_y,-k_x,0),
\end{equation}
with
\begin{equation}
    \aR=\frac{2\tilde a\tilde b}{\Eg-b}.
\end{equation}
The second term in Eq.~(\ref{H-eff-pp-gen}) is the effective antisymmetric SOC generated in the $\Gamma_1$ orbital manifold by its coupling with the $\Gamma_5$ orbital. 
The spin-current operators have the form 
\begin{widetext}
\begin{multline}
    \hat{\cal J}_{\mu,i}(\bk) = \left[1-\frac{2m\Eg\tilde a^2}{\hbar^2(\Eg^2-b^2)}-\frac{2\tilde b^2}{(\Eg-b)^2}-\frac{2b^2\tilde a^2k^2}{(\Eg^2-b^2)^2}\right]\frac{\hbar k_i}{m}\hat\sigma_x
    -\frac{2b\tilde a^2}{\hbar(\Eg^2-b^2)}(\delta_{ix}k_y-\delta_{iy}k_x)(\delta_{\mu x}\hat\sigma_y-\delta_{\mu y}\hat\sigma_x) \\
     +\frac{2\tilde a\tilde b}{\hbar(\Eg-b)}\left[\delta_{\mu x}\delta_{iy}-\delta_{\mu y}\delta_{ix}+\frac{2b}{\Eg^2-b^2}\frac{\hbar^2k_i}{m}(\delta_{\mu x}k_y-\delta_{\mu y}k_x)\right]\hat\sigma_0
    \end{multline}
for $\mu=x,y$, and
\begin{equation}
    \hat{\cal J}_{z,i}(\bk) = \left[1-\frac{2m\Eg\tilde a^2}{\hbar^2(\Eg^2-b^2)}-\frac{4\tilde b^2}{(\Eg-b)^2}\right]\frac{\hbar k_i}{m}\hat\sigma_z,
\end{equation}
while the effective spin operators are given by
\begin{eqnarray*}
  \hat{\cal S}_\mu(\bk) &=& \left[1-\frac{2\tilde b^2}{(\Eg-b)^2}-\frac{2b^2\tilde a^2k^2}{(\Eg^2-b^2)^2}\right]\hat\sigma_\mu
      +\frac{4b\tilde a\tilde b}{(\Eg-b)(\Eg^2-b^2)}(\delta_{\mu x}k_y-\delta_{\mu y}k_x)\hat\sigma_0,\quad \mu=x,y, \\
  \hat{\cal S}_{z}(\bk) &=& \left[1-\frac{4\tilde b^2}{(\Eg-b)^2}\right]\hat\sigma_z.
\end{eqnarray*}
\end{widetext}
For the effective spin torque $\hat{\cal T}_\mu=(i/\hbar)[\hcH,\hat{\cal S}_\mu]$ we obtain:
\begin{eqnarray*}
    \hat{\cal T}_\mu(\bk) &=& -\frac{2\aR}{\hbar}k_\mu\hat\sigma_z,\quad \mu=x,y, \\
    \hat{\cal T}_z(\bk) &=& \frac{2\aR}{\hbar}(k_x\hat\sigma_x+k_y\hat\sigma_y). \nonumber
\end{eqnarray*}
The above expressions are valid to second order in the inter-orbital couplings, i.e. in $\tilde a$ and $\tilde b$, while the intra-orbital SOC in the 2D orbital ($b$) is taken into account exactly.
Introducing the renormalized parameter $\tilde b(1+b/\Eg)^{-1}\to\tilde b$ and expanding in powers of $b$ at $b\ll\Eg$, we arrive at Eq.~(\ref{J-mu-i-dd-final}).

%==================================================================
%==================================================================

\section{Spin current in an isolated orbital}
\label{app: isolated orbital}

The simplest model of a band structure includes just one isolated orbital state corresponding to one of the irreps of $\mathbb{G}=\mathbf{C}_{4v}$, with all inter-orbital transitions neglected. In this case, there is no need to use the effective Hamiltonian formalism and the spin current can be calculated exactly. In particular, for an isolated 1D orbital the equilibrium spin current was obtained in Appendix \ref{app: trivial-topological-toy}:
\begin{equation}
\label{j-xy-intrinsic-1D}
  J_s(T=0)=\frac{m^{2}}{3\pi\hbar^5}\alpha_0^3,
\end{equation}
where $\alpha_0$ is the strength of the intrinsic Rashba SOC. 
 
For one isolated 2D $\Gamma_5$ orbital, the microscopic Hamiltonian is obtained by truncating Eq.~(\ref{H-p-p-general}) or (\ref{H-d-d-general}) to the lower-right $4\times 4$ block:
\begin{eqnarray}
\label{H-d-isolated-2D}
  \hat{\msH}(\bk) = \hat\tau_0\otimes\left[\frac{\hbar^2k^2}{2m}\hat\sigma_0+\alpha_0(k_y\hat\sigma_x-k_x\hat\sigma_y)\right] 
  \nonumber\\
  +b(\hat\tau_2\otimes\hat\sigma_z),
\end{eqnarray}
where $\alpha_0=\gamma_2$ is the strength of the intrinsic Rashba SOC, $b$ corresponds to the local intra-orbital SOC, and the energy is counted from $\epsilon_2$. This Hamiltonian can be diagonalized as follows:
$$
  \hat{\msW}^{-1}(\bk)\hat{\msH}(\bk)\hat{\msW}(\bk)=\mathrm{diag}\left[\xi_+(\bk),\xi_-(\bk),\xi_+(\bk),\xi_-(\bk)\right],
$$
where 
\begin{equation}
\label{isolated-2D-bands}
  \xi_\pm(\bk)=\frac{\hbar^2k^2}{2m}\pm\sqrt{\alpha_0^2k^2+b^2}.
\end{equation}
Introducing the notation $\bm{\Gamma}_\pm(\bk)=(\alpha_0k_y,-\alpha_0k_x,\pm b)$, we have
\begin{equation}
\label{isolated-2D-W}
  \hat{\msW}=\frac{1}{\sqrt{2}}\left( \begin{array}{cc}
                                  \hat U_+ & i\hat U_- \\
                                  i\hat U_+ & \hat U_-
                                  \end{array} \right),
\end{equation}
where 
$$
  \hat U_\pm=\left( \begin{array}{cc}
                 \cos\dfrac{\theta_\pm}{2} & e^{-i\varphi_\pm}\sin\dfrac{\theta_\pm}{2} \\ 
                    e^{i\varphi_\pm}\sin\dfrac{\theta_\pm}{2} & -\cos\dfrac{\theta_\pm}{2}
                    \end{array} \right)
$$
are the matrices diagonalizing $\bm{\Gamma}_\pm\cdot\hbs$ and we used the angular parameterization (\ref{gamma-parameter}), with 
$\tan\theta_+=-\tan\theta_-=\alpha_0k/b$ and $\varphi_+=\varphi_-=\arg\bk-\pi/2$. Note that the bands remain twofold degenerate at each $\bk$, despite the absence of inversion symmetry, because the truncated 
Hamiltonian (\ref{H-d-isolated-2D}) commutes with $\hat\tau_2$. When the coupling of the $\Gamma_5$ orbital with other orbitals is taken into account the degeneracy will be lifted, except at the TR invariant points, as shown in Fig. \ref{fig: G1G5 bands}.

Applying the definition (\ref{J-mu-i-exact}) to the Hamiltonian (\ref{H-d-isolated-2D}), we obtain the following microscopic spin~current operator in an isolated $\Gamma_5$ orbital:
$$
   \hat{\msJ}_{\mu,i}(\bk)=\hat\tau_0\otimes\left[\frac{\hbar k_i}{m}\hat\sigma_\mu+\frac{\alpha_0}{\hbar}(\delta_{\mu x}\delta_{iy}-\delta_{\mu y}\delta_{ix})\hat\sigma_0\right].
$$
This can be transformed into the band representation,
\begin{eqnarray*}
    \hat{\msW}^{-1}\hat{\msJ}_{\mu,i}\hat{\msW}=\frac{\hbar k_i}{m} \left( \begin{array}{cc}
            \hat U_+^{-1}\hat\sigma_\mu\hat U_+ & 0 \\
            0 & \hat U_-^{-1}\hat\sigma_\mu\hat U_-
            \end{array} \right) \\
  +\frac{\alpha_0}{\hbar}(\delta_{\mu x}\delta_{iy}-\delta_{\mu y}\delta_{ix})(\hat\tau_0\otimes\hat\sigma_0),
\end{eqnarray*}
from which we obtain the spin-current operators in the 2D subspaces of the bands (\ref{isolated-2D-bands}),
\begin{eqnarray}
\label{isolated-2D-spin-current}
    J^\pm_{\mu,i}(\bk) &=& \pm\frac{\hbar k_i}{m}\frac{1}{\sqrt{\alpha_0^2k^2+b^2} } \left( \begin{array}{cc}
        \Gamma_{+,\mu}(\bk) & 0 \\
        0 & \Gamma_{-,\mu}(\bk)
        \end{array} \right) \nonumber \\
    && +\frac{\alpha_0}{\hbar}(\delta_{\mu x}\delta_{iy}-\delta_{\mu y}\delta_{ix})\hat\sigma_0.
\end{eqnarray}

The equilibrium spin current is given by
$$
  \langle j_{\mu,i}\rangle = \int\frac{d^2\bk}{(2\pi)^2} \left[\tr(J^+_{\mu,i}) f(\xi_+)+\tr(J^-_{\mu,i}) f(\xi_-)\right].
$$
From Eq.~(\ref{isolated-2D-spin-current}) it is easy to see that the only nonzero components are $\langle j_{x,y}\rangle=-\langle j_{y,x}\rangle=J_s(T)$. At zero temperature, the calculation is straightforward and we obtain:
\begin{equation}
\label{j-xy-intrinsic-2D}
  J_s(T=0)=\frac{2m^{2}}{3\pi\hbar^5}\alpha_0^3,
\end{equation}
where we put $\mu>|b|$, so that both bands (\ref{isolated-2D-bands}) cross the chemical potential.
The last expression exhibits the same features as Eq. (\ref{j-xy-intrinsic-1D}), namely, it is cubic in the strength of the intrinsic Rashba SOC and independent of the chemical potential. The additional factor of two can be attributed to the double degeneracy of the bands. 
Notably, the equilibrium spin current in an isolated $\Gamma_5$ orbital state does not depend on $b$, i.e. on the local intra-orbital SOC.

%==================================================================
%==================================================================

%\bibliography{spin-currents}

\begin{thebibliography}{99}
\bibitem{spin-current-units}
Following the considerable precedent, see, e.g. Ref.~\cite{Rashba03}, we use the dimensionless Pauli matrices instead of the spin angular momentum operators $(\hbar/2)\hbs$ in all definitions of the spin observables.
\bibitem{2D-NCS-meaning}
In a 2D noncentrosymmetric crystal, the lattice potential $U(x,y,z)$ is periodic in $x$ and $y$, but confining in $z$. The mirror reflection symmetry $z\to-z$ is absent, i.e. $U(x,y,z)\neq U(x,y,-z)$.
\bibitem{2D-orbital-isotropy}
For the sake of simpler expressions, we represent the dispersion of $\varepsilon_2(\bk)$ as isotropic, which corresponds to the situation where all other orbitals are infinitely remote in energy. This approximation has no influence on our conclusions.
\bibitem{no-ext-E}
In some systems, e.g. in semiconductor heterostructures, inversion symmetry can also be broken by an applied electric field. Our theory can be straightforwardly extended to such cases.
\end{thebibliography}

\end{document}